\pdfoutput=1
\documentclass[usenatbib]{mn2e}
\title[Dark matter in disk galaxies I]{Dark matter in disk galaxies I: a Markov Chain Monte Carlo method and application to DDO 154}
\author[Hague, P. R., Wilkinson, M. I.]{
P. R. Hague~\thanks{prh19@le.ac.uk} \&
M.I. Wilkinson~\thanks{miw6@le.ac.uk}\\
Department of Physics \& Astronomy, University of Leicester, 
  University Road, Leicester LE1 7RH, UK}

\usepackage{graphicx}
\usepackage{subfig}
\usepackage{color}
\usepackage{amssymb,amsmath}

\newcommand{\kms}{km\,s$^{-1}$}

\newcommand{\imgext}{png}

\begin{document}

\maketitle
\begin{abstract}
We present a new method to constrain the dark matter halo density profiles of disk galaxies. Our algorithm employs a Markov Chain Monte Carlo (MCMC) approach to explore the parameter space of a general family of dark matter profiles. We improve upon previous analyses by considering a wider range of halo profiles and by explicitly identifying cases in which the data are insufficient to break the degeneracies between the model parameters.  We demonstrate the robustness of our algorithm using artificial data sets and show that reliable estimates of the halo density profile can be obtained from data of comparable quality to those currently available for low surface brightness (LSB) galaxies. We present our results in terms of physical quantities which are constrained by the data, and find that the logarithmic slope of the halo density profile at the radius of the innermost data point of a measured rotation curve can be strongly constrained in LSB ([$v_{\rm star} / v_{\rm obs}]_{\rm max} \simeq 0.16$) galaxies. High surface brightness galaxies ([$v_{\rm star} / v_{\rm obs}]_{\rm max} \simeq 0.79$)  require additional information on the mass-to-light ratio of the stellar population - our approach naturally identifies those galaxies for which this is necessary.

We apply our method to observed data for the dwarf irregular galaxy DDO 154 and recover a logarithmic halo slope of $-0.39 \pm 0.11$ at a radius of 0.14 kpc. Our analysis validates earlier estimates which were based on the fitting of a limited set of individual halo models, but constitutes a more robust constraint than was possible using other techniques since it marginalises over a wide range of halo profiles. Our method can thus reproduce existing results, has been verified on test data, and is shown to be capable of providing more information than is available from fitting individual halo profiles. The likely impact of future improvements in data quality on rotation curve decomposition using this technique is also discussed. We find that velocity errors are a limiting factor on the constraint that can be found, while spatial resolution is not.
\end{abstract}


\section{Introduction}

In order to understand the process of galaxy formation, it is important to have robust constraints on the gravitational potential wells in which observed galaxies reside. According to the current $\Lambda$ Cold Dark Matter ($\Lambda$CDM) cosmological paradigm, dark matter haloes are generally thought to be an essential requirement for formation of a galaxy as they provide the means to collect and bind sufficient baryonic matter to create galaxies of sizes consistent with observations. Measurements of the velocities of gas moving in the disks of disk galaxies provide a valuable probe of the total gravitational potentials of such systems. Pioneering work by~\cite{bosma1978} and~\cite{rubin1978} demonstrated that the rotation curves of disk galaxies were generally flat to large radii, and the observed disparity between the measured rotation curves and those predicted based on their luminous components is now generally interpreted as evidence for the presence of dark matter haloes in these galaxies.  

Observations of such a galaxy can give the rotation velocity at a specific radius $r$ from the centre of the galaxy, the density of gas at this distance, and the amount of stellar light (and, hence the stellar mass, modulo certain assumptions about the mass-to-light ratio $\Upsilon$) at each $r$. Calculating the rotation curves that would be expected from the two baryonic components generally gives a combined value that falls short of the observed rotation curve, and the difference can be used to infer the amount of dark matter enclosed at each $r$ and thus a density profile for the halo.

While the ubiquity of dark matter haloes around galaxies is by now well established, observational determinations of the density \textit{profiles} of galactic dark matter haloes have been, and remain, controversial.  \cite{flores1994} used gravitational lensing and rotation curve analysis to argue against cusped\footnote{In the literature, the terms cusp and core generally refer to negative log slopes at $r=0$ of $\gamma_0 \geqslant 1$ and $\gamma_0=0$ respectively. In this paper we discuss halos with intermediate inner slopes. We considered such slopes neither cusped nor cored, even though in the strict mathematical sense any halo $\gamma_0 > 0$ is cusped, in order to maintain coherence with previous discussions.} haloes with $\rho(r)\sim r^{-1}$ and $\rho(r)\sim r^{-2}$, which were being predicted by N-body simulations of cosmological structure formation at that time~\citep[e.g. ][who found that the profile proposed by \cite{hernquist1990} best fit the haloes they obtained]{dubinski1991}. During the 1990s, dark matter-only cosmological simulations continued to suggest that a universal, cusped dark matter density profile, independent of halo mass, should be the outcome of the hierarchical formation scenario for galaxy haloes~\citep[][hereafter NFW]{navarro1996,navarro1997}. On the other hand, a number of authors argued that observations of disk galaxy rotation curves pointed towards the existence of a universal rotation curve~\citep{ps1991,burkert1995,pss1996}. The former claimed an inner halo profile of  $\rho(r)\sim r^{-1}$ whilst the latter found uniform density central cores, and stated that the cusps implied by numerical simulations could be excluded. The apparent disparity between the results of cosmological simulations and observations of real galaxies became known as the ``cusp-core controversy". 

Early work to resolve this controversy focused on low surface brightness (LSB) galaxies, on the basis that the stellar contribution could be ignored entirely in a first-order model. In \cite{vandenbosch2001} it was claimed that it was impossible to differentiate between flat cores and $r^{-1}$ cusps using the data available at the time. They cited the insufficient radial range of the data, and the problem of beam smearing. Beam smearing of HI rotation curve data is caused by the fact that the each beam is larger than the scale over which the rotation speed changes. This, and several other observational issues that could lead to the false inference of cored haloes in galaxies which actually contained dark matter cusps, were investigated and discounted by \cite{deblok2003}. Using direct inversion of a large sample of rotation curves to determine mass profiles, they also found that the outer regions of their sample galaxies were consistent with density profiles $\rho\sim r^{-2}$ while the inner regions required power laws which were typically shallower than $\rho\sim r^{-0.25}$.  

\cite{bosma2003} claimed to rule out $r^{-1}$ cusps at the 3-sigma confidence level for 17 galaxies (from a sample of 28) and, on this basis, stated that the slopes predicted by CDM models were Ônot observedÕ. He demonstrated that the position of the slit used at the time to observe rotation curves could not have been a factor in this result, by deliberately offsetting the slit and observing the effect on the measured rotation curve. ~\cite{gentile2004} subsequently found that the cored Burkert profile provided higher quality fits to the rotation curves of five nearby galaxies than either the cusped Navarro-Frenk-White (NFW) or a model based on Modified Newtonian Dynamics (MOND) that would not require dark matter. This growing weight of evidence has led to the widespread acceptance of the observational existence of cores. However, it is now recognised that exclusion of cusps in present-day galaxies is not synonymous with invalidating the $\Lambda$CDM paradigm. The profile of a galactic halo results from the interplay between a number of factors including the intrinsic physical properties of dark matter, the re-distribution of baryonic material by gas cooling, supernovae or AGN feedback, and its history of major or minor mergers with other galaxies~\citep[see e.g.][for recent simulations of these processes at work on galaxy scales]{Governato2012}. Cored haloes may originally have had cusps - observations of the present day profiles constrain the end-state of galaxy formation, rather than the initial conditions.

In this context, it is important to note that ``measured'' halo properties should not be confused with the values of particular parameters in a profile with an assumed form. For example, a halo with an asymptotic inner slope of zero may exhibit a non-zero slope over the entire radial range probed by a particular observational data set and indeed may be well reproduced by an NFW profile over that radial range for an appropriate choice of parameters. Additional information may be used to argue that the NFW parameters are unrealistic (for example, based on their expected values using the scaling relations obtained from cosmological simulations).  In our analysis, we therefore focus on physically meaningful quantities such as the logarithmic slope at the radius of the innermost point in the rotation curve, rather than the values of particular parameters in our models.

The HI Nearby Galaxy Survey (THINGS) sample of galaxies provides high resolution 2D data which addresses many of the issues such as non-circular motions that have plagued rotation curve decompositions in the past~\citep{THINGS}. It is therefore timely to examine in detail the constraints which can be placed on the physical properties of the haloes of disk galaxies using these high resolution observations of their rotation curves. Somewhat surprisingly, research to date has generally focussed on specific halo models~\citep[e.g.][]{deblok2008}, with particular interest in the NFW and Burkert profiles (see Sec.~\ref{sec:models}). Recently, \cite{chemin2011} re-analysed the THINGS rotation curves using Einasto profiles and found that these provided improved fits to the observed data relative to earlier work. This is to be expected, as the Einasto profiles have an additional shape parameter which makes their form more flexible for the modelling of rotation curves.

Taking advantage of new data from THINGS, \cite{oh2011} claim that a selection of dwarf galaxies (including DDO 154, the galaxy that we use as an example in this paper) exhibit $r^{-0.29}$ inner profiles, which they classify as cores. This conclusion assumes a single dark matter halo for all dwarf galaxies, but not necessarily a truly universal profile. DDO 154 is a common target for studies of dark matter because it has a low surface brightness, and now thanks to THINGS there is more extensive rotation curve data for this galaxy than was available in older studies~\citep{THINGS}.

All the above analyses focus on a small number of possible profiles, and thus any comparison between them can only reveal which of the profiles considered best fits the data. They provide no context in which to discuss how well the ``best'' profile performs relative to all possible profiles. Furthermore, by restricting the set of profiles considered, these earlier studies have to some extent dictated the shape of the model rotation curve, so the slope of the best-fit model curve at any individual radius is based on the fit statistics of the entire curve. It is possible that the quality of the fit for one part of the curve might give a false impression of the quality of the fit for another part of the curve, when in actual fact there is not enough information at that point to constrain the slope.

In this paper, we re-visit the problem of rotation curve decomposition and use a Markov Chain Monte Carlo (MCMC) algorithm to explore the parameter space of a very general family of dark matter haloes. ~\cite{puglielli2010} have previously applied MCMC techniques to the problem of rotation curve decomposition, but considered a restricted set of halo profiles. Our three key innovations are: (1) the use of a more general family of halo profiles than has previously been used, thereby permitting the models more freedom to match the observed data; (2) a focus on physical quantities which can be constrained by the data, for example, the log-slope of the halo profile at the innermost observed data point rather than the particular parameters of our models, for example the asymptotic values of the log-slope of the dark halo profile at radii smaller or larger than those probed by the data; (3) the potential for a detailed exploration of the degeneracies between the model parameters. This latter exploration enables us both to quantify the true uncertainties in the halo constraints we obtain as well as to determine what future observations would be most likely to improve these constraints. 

The outline of the paper is as follows. Section~\ref{sec:models} describes the galaxy models we use, both to analyse our artificially-generated data with known input values, and in our analysis of the THINGS data for DDO 154 obtained from~\cite{deblok2008}. In Section~\ref{sec:algorithm} we present the algorithm we have developed for this purpose, employing CosmoMC ~\citep{lewis2002} as an MCMC driver. Section~\ref{sec:tests} details the results of tests on artificial data while Section~\ref{sec:comparison} compares the performance of our method with that of existing rotation curve fitting techniques (i.e. fits based on individual or small numbers of dark matter halo profiles). In Section~\ref{sec:ddo154} we present an application of our method to DDO 154 and finally Section~\ref{sec:conc} details our conclusions. An appendix is included which presents some additional technical details of our method.

\section{Galaxy Models}
\label{sec:models}

Under the assumption that the gas in a disk galaxy disk moves along approximately circular orbits, the measured run of gas velocity with radius through the disk is a strong probe of the total underlying mass distribution. There are three main contributions to the mass distribution which we must account for in any mass model: the stellar disk, the gas disk, and the dark matter halo. The spatial distributions of these three components are significantly different and therefore their relative contributions are strong functions of radius. In the inner regions of many of the THINGS galaxies, the baryonic components contribute significantly to the gravitating mass, making it more difficult to determine the inner slope of the dark matter profile. In this paper, we go beyond previous analyses by considering a more general family of halo density profiles and by explicitly exploring degeneracies between the parameters of our models, in particular between the dark matter and baryonic parameters. 

\subsection{Baryonic components}

The Spitzer Infrared Nearby Galaxies Survey~\citep[SINGS: see][]{kennicutt2003} provides near-infrared photometric data for the galaxies in the THINGS catalogue, allowing the determination of accurate stellar surface density profiles. Conversion of the luminosity profile to a stellar mass profile requires information about the mass-to-light ratio of the stellar population. In this paper, we make use of the stellar rotation curves calculated by \cite{deblok2008}. These curves are calculated by assuming that radial variations in the colour of the stellar populations indicate stellar population gradients which in turn lead to variations in $\Upsilon_{\rm 3.6}$, the mass-to-light ratio in the $3.6\mu$m band. As described below, our algorithm incorporates a scaling of the stellar rotation curve to account for uncertainties in the stellar population modelling. The same scaling is applied at all radii, and we thus implicitly assume the same radial gradients in $\Upsilon$ as de Blok et al. (2008).

In Section~\ref{sec:tests}, we use DDO 154 as a template for LSB galaxies and NGC 7793 as an example of an HSB galaxy.
We use the data for their stellar disks from \cite{deblok2008}. The surface brightness measured in the $3.6\rm{\mu m}$ band for NGC 7793 was taken from SINGS, while that for DDO 154 was obtained from the Spitzer archive. A $\Upsilon$ value was then calculated from the $J - K$ band of 2MASS, using a formula that assumes a ``diet" Salpeter IMF (one with fewer low mass stars, necessary to keep some stellar disks sub-maximal). \cite{deblok2008} also present results using a Kroupa IMF. However as we anticipate most difficulty analysing HSB galaxies, we choose the option that gives a more massive disk in order to find the limit of the effectiveness of our method. 

\cite{deblok2008} combined the surface brightness and $\Upsilon$ into a density profile, which they then converted to a velocity curve using the GIPSY\footnote{http://www.astro.rug.nl/$\sim$gipsy/} software. We approximate this velocity curve by fitting a thin exponential disk rotation curve~\citep[see e.g.][]{BT} to produce an estimate of the stellar contribution to the total rotation curve. The goodness of fit for this profile does not impact on the performance of our algorithm, as we marginalise over both the scale radius $R_{\rm d}$ and the amplitude (by varying the stellar $\Upsilon$: see below). However, we note that it provides an excellent fit to the data for DDO 154, with a maximal deviation of $\lesssim0.25$\kms over the entire radial range, and significantly less than this interior to 1.5kpc. 

In our analysis, we apply a scaling factor $f_{\Upsilon}$ to explore the impact of observational uncertainty on the mass of the stellar disk. The inclusion of $f_{\Upsilon}$ as a free parameter in our analysis enables us to explore whether we can differentiate between those cases in which there is sufficient information to constrain the halo profile despite our lack of knowledge about the true $\Upsilon$ and those in which there is not. The latter cases will necessarily be beyond the scope of this technique until $\Upsilon$ can be determined with greater accuracy. However, by reducing the range of allowed $f_{\Upsilon}$ values explored by our algorithm, we would be able to determine what $\Upsilon$ precision is required to constrain the halo to a particular level of certainty. We allow $f_{\Upsilon}$ to vary between 0.1 and 2, which encompasses $\Upsilon$ values derived from the full Salpeter IMF ($f_{\Upsilon}$ = 1.43) and the Kroupa IMF ($f_{\Upsilon}$ = 0.71) that have been considered in the previous analyses of rotation curve by \cite{chemin2011} and \cite{deblok2008}, without placing either one near the edge of the parameter space.

Radial variations in $\Upsilon$, which occur due to age gradients in the disk and which are detected by means of color gradients, are already taken into account in the stellar rotation curves of \cite{deblok2008} which we use in our modelling. This, and the fact that the $3.6{\rm \mu m}$ band is sensitive mainly to the older portion of the stellar population, means that our modelling technique is robust with respect to the stellar age.

The gas distribution for the THINGS galaxies is determined from their HI maps. The gas mass is estimated assuming that the gas is entirely composed of atomic hydrogen. While the presence of molecular gas would change the gas contribution to the overall potential well, we note that for the majority of galaxies in the THINGS survey, the ISM is dominated by atomic hydrogen~\citep{leroy2008}. However, for galaxies with a significant gas contribution to the gravitating mass, the composition of the ISM should be confirmed to test this assumption.

Similarly to the the stellar disk, we take our model gas profiles from the \cite{deblok2008} data for DDO 154 and NGC 7793. Following \cite{deblok2008}, we have scaled the surface density given by THINGS HI data cubes by a factor of 1.4  in order to take into account the non-hydrogen contributions of the gas (galaxies with large amount of molecular gas having already been excluded from analysis at this point). They then used the GIPSY software to construct an infinitely thin disk rotation curve based on a titled ring modelling of the gas disk. We take this profile and apply the same smoothing to it as we did for the stellar curve to obtain the gas contribution to the rotation curve. No variation of the parameters of the gas disk are allowed in our modelling, as the mass and extent of the disk are assumed to be well-constrained by the observations.

In addition to the gas located in the disk, galaxies may also have a hot, ionised halo of gas. \cite{miller2013} find that, in the Milky Way, the fraction of the total galaxy mass in this component is 0.07. This means that, assuming comparable fractions are present in nearby disk galaxies, the kinematic effect of this halo is smaller than errors due to non-circular motions.

It is well known that non-circular motions complicate the interpretation of galaxy rotation curves because the observed gas motions at a particular radius may not accurately reflect the underlying circular velocity at that location. \cite{pizzella2008} show that rotation curves based on stellar velocities may be more reliable for determinations of the shape of the inner halo of a galaxy than those based on one-dimensional HI spectra. They note, however, that integral field gas velocity maps can be used to screen out galaxies with significant non-circular contributions to their velocity fields. The THINGS sample has been selected on the basis of strict criteria including favourable inclination angle relative to the line of sight~\citep{deblok2008}. Subsequently, \cite{oh2011} examined the velocity field data for evidence of non-circular motions and determined an optimal rotation curve for each THINGS galaxy, which thus represents the best opportunity to determine the dark halo profiles of this sample. In the present paper, we therefore assume that the galaxy to be modelled has been checked for non-circular motion, and that the rotation curve has been corrected for such motions if required.

A further potential complication is asymmetric drift, which affects measurements of circular speeds obtained from gas or stars whose vertical velocity dispersion is not negligible compared to their ordered rotational velocities. In our modelling, we assume that asymmetric drift can be neglected, or has been corrected for in the rotation curve data, and as in \cite{deblok2008} assume an infinitely thin gas disk.

\subsection{Halo models}

In common with most previous analyses, we assume that the dark matter halo of the galaxy is spherical. Whilst non-spherical haloes are in principle possible, and indeed are the likely outcome of either mergers and/or gas cooling~\citep[generically leading to triaxial haloes; see e.g.][for a discussion]{veraciro2011}, the presence of triaxiality would typically lead to non-circular motion of the stars and gas in the disk. In their analysis of the rotation curve of DDO 47, ~\cite{gentile2005} argued that the non-circular motions seen in that galaxy were at a level of $\lesssim 3$\kms and concluded that any triaxiality in the halo of that galaxy was at a level which was irrelevant for rotation curve analyses. A study of the non-circular motions in the THINGS galaxies by \cite{trachternach2008}, based on harmonic decomposition of the velocity around each tilted ring, found that the amplitude of non-circular components was less than 1 km ${\rm s}^{-1}$ at all radii. Given that we are restricting our analysis to such galaxies which exhibit very low levels of non-circular motion, the assumption of a spherical halo is reasonable. 

We assume that the dark matter halo can be parameterised by an ($\alpha,\beta,\gamma$) profile~\citep{zhao1996}, which is a general, spherical halo density profile given by

\begin{equation}
\rho(r) = \frac{\rho_{\rm s}}{\displaystyle\left(\frac{r}{r_{\rm s}}\right)^\gamma \displaystyle\left(1+\left(\frac{r}{r_{\rm s}}\right)^{1/\alpha}\right)^{\alpha ( \beta - \gamma)}} 
\end{equation}

where $\gamma$ is the asymptotic log-slope of the profile at small $r$, $\beta$ is the slope at large $r$, and $\alpha$ controls the transition between the two (lower values of $\alpha$ corresponding to sharper changes).  The parameters $\rho_{\rm s}$ and $r_{\rm s}$ are the density and length scales, respectively. This family encompasses a number of the profiles which have previously been used to model disk galaxy haloes. The cusped NFW profile, proposed as a good approximation to the dark matter halo profiles obtained in dark mater-only cosmological simulations of structure formation, corresponds to $(\alpha,\beta,\gamma) = (1,3,1)$, while the cored isothermal halo, which is often used as an example of a cored halo profile has $(\alpha,\beta,\gamma) = (1/2,2,0)$. Another widely used cored halo is the~\cite{burkert1995} profile,
\begin{equation}
\rho(r) = \frac{\rho_{\rm s} r_{\rm s}^3}{({r + r_{\rm s}})(r^2 + r_{\rm s}^2)} 
\end{equation}
which has been shown to provide a good fit to the rotation curves of a large sample of disk galaxy rotation curves~\citep[see e.g.][and references therein]{ps1988,ps1991,salucci2007}. More recently, \cite{chemin2011} showed that the Einasto profile~\citep{einasto1969,einasto1965},
\begin{equation}
\rho(r)=\rho_s \exp\left[{-\frac{2}{n}\left[\left(\frac{r}{r_{\rm s}}\right)^n-1\right]}\right]
\end{equation}
where $r_{\rm s}$ is a scale radius, $\rho_{\rm s}$ is a density scale and $n$ is a shape parameter, provides a better fit to the rotation curves of THINGS galaxies than the other three profiles. This profile, whose logarithmic slope varies continuously with radius, has been proposed as an improvement on the NFW profile in terms of fitting the results of the most recent cosmological simulations~\citep{navarro2004}. While the ($\alpha,\beta,\gamma$) family does not explicitly include either the Burkert or Einasto profile, for suitable choices of the profile parameters it can closely reproduce them over a restricted radial range. 

For any ($\alpha,\beta,\gamma$) model, the enclosed mass at a particular radius is
\begin{equation}
M={4 \pi \rho_{\rm s} r^3 \over \gamma-3 } \left({r \over r_{\rm s}}\right)^{-\gamma}  {_2 F_1} (a,b,c,z)
\end{equation}
where $_2 F _1()$ is the hypergeometric function, with $a=\alpha[\beta-\gamma]$, $
b=-\alpha[\gamma-3]$, $c=1-\alpha[\gamma-3]$, and $z=-[\frac{r}{r_{\rm s}}]^{1/\alpha}$.

In this paper, we re-cast the ($\alpha,\beta,\gamma$) profile in the form
\begin{equation}
\rho(r) = \frac{\Sigma_{\rm max}}{G} \frac{v_{\rm{max}}^2}{\displaystyle \left(\frac{r}{r_s}\right)^\gamma \left(1+\left[\frac{r}{r_s}\right]^{1/\alpha}\right)^{\alpha ( \beta - \gamma)}} 
\end{equation}
with $v_{\rm max}$ replacing the $\rho_{\rm s}$ paramater, and $\Sigma_{\rm{max}}$ being calculated from the remaining parameters via the formula
\begin{equation}
\Sigma_{\rm max} = \frac{\rho_{\rm s}r_{\rm max}}{M(\alpha, \beta, \gamma, r_{\rm s}, \rho_{\rm s})}
\end{equation}
This transformation is explained in detail in Appendix~\ref{paramtrans}. Note that this is a different parameterisation of the same halo, rather than a distinct halo itself. The reason for introducing this parameter transform is to resolve the degeneracy between the halo parameters $\rho_{\rm s}$ and $r_{\rm s}$. The parameter $v_{\rm max}$ is a useful choice as it has a clear physical meaning, namely, the maximum circular velocity of the dark matter halo. This should not be confused with the maximum circular velocity of the observed rotation curve, although in the case of LSB galaxies they will be similar in value.

\section{Markov Chain Monte Carlo}
\label{sec:algorithm}

\begin{table}
\begin{tabular}{lcc}
  Parameter & Minimum & Maximum \\ \hline
  $\alpha$ & 0.1 & 2.5   \\
  $\beta$ & 3 & 5  \\
  $\gamma$ & 0 & 2  \\
  $r_{\rm s}$ & $r_{\rm min}$ & $r_{\rm min}+2r_{\rm s, 0}$  \\
  $v_{\rm max}$ & 0 & $2v_{\rm max, 0}$  \\
  $f_{\Upsilon}$ & 0.1 & 2  \\
  $f_{R_{\rm d}}$ & 0.1 & 2  \\ \hline
\end{tabular}
\caption{Summary of the parameters explored by the algorithm. $r_{\rm s, 0}$ and $v_{\rm max, 0}$ are initial values determined by a single $\chi^2$ best fit. $r_{\rm min}$ is the radius of the smallest radial bin of a particular data set. $f_{R_{\rm d}}$ is a scaling factor which is used to vary the stellar disk scale length.}
\label{tblparams}
\end{table}

\begin{table*}
\begin{tabular}{lccccccccl}
 Model & $\alpha$ & $\beta$ & $\gamma$ & $\gamma_{\rm in}$ & $r_{\rm s}$ & $v_{\rm max}$ & $R_{\rm d}$ & ${\Delta v \over v}$ & Notes \\ \hline
A &1 & 4 & 1 & 1.07 & 6 & 50 & 2.15 & 0.02 & Low surface brightness with small errors  \\
B &1 & 4 & 1 & 1.07 & 6 & 50 & 2.15 & 0.1 & Low surface brightness with realistic errors  \\
C &1 & 4 & 1 & 1.05 & 6 & 80 & 2.50 & 0.1 & High surface brightness with realistic errors$^1$ \\
D &1 & 4 & 1 & 1.07 & 6 & 50 & 2.15 & 0.02 & Low surface brightness with high data resolution \\
E &1 & 4 & 1 & 1.07 & 6 & 50 &  2.15 & 0.1 & Low surface brightness with realistic errors, \\
&&&&&&&&& extended radial range \\
F & $-$ & $-$ & $-$ & 0.08 & 6 & 50 &  2.15 & 0.1 & Low surface brightness with realistic errors, \\
&&&&&&&&& Burkert profile$^2$ \\
G &1 & 4 & 1 & 1.07 & 6 & 50 &  2.15 & 0.1 & Low surface brightness with realistic errors,  \\
&&&&&&&&& free stellar disk parameters \\
H &1 & 4 & 1 & 1.05 & 6 & 80 &  2.50 & 0.1 & High surface brightness with realistic errors,  \\
&&&&&&&&& free stellar disk parameters$^1$ \\
 \end{tabular}
   \caption{Input parameters for synthetic data sets. The columns show: (1) Model name; (2-4) ($\alpha,\beta,\gamma$) parameters of the halo profile; (5) log slope of the halo profile at the innermost data point, $\gamma_{\rm in}$; (6) halo scale radius in kpc; (7) maximum velocity of the halo circular speed curve in \kms; (8)  disk scale length in kpc; (9) observational error. All except Model F use ($\alpha,\beta,\gamma$) profile haloes, with parameter values corresponding to a~\protect\cite{hernquist1990} halo. All models except G and H assume a $f_{\Upsilon}$ of $1$. For the LSB models, the innermost data point is located at a radius of $0.14$kpc (based on DDO 154) and for the HSB models, the innermost data point is at a radius of $0.11$kpc (based on NGC 7793). Notes: $^1$ the $\gamma_{\rm in}$ values for Model C and Model H are not directly comparable with the rest of the values as the radius at which they are measured is different; $^2$ the scale radius and scale density in the Burkert profile is not directly comparable to those from the ($\alpha,\beta,\gamma$) profiles. }
   \label{tab:testparams}
\end{table*}

Our goal is to determine the distributions of parameters for the models described in the previous section which are consistent with a given observed HI rotation curve. We assume that the observed rotation curve can be modelled by summing, in quadrature, the contributions to the local circular speed of the gas disk, the stellar disk and the dark matter halo. For a given choice of model parameters, we calculate the expected rotation curve which we compare to the observed data by means of a $\chi^2$ test.

Markov Chain Monte Carlo~\citep[MCMC; see e.g.][]{NR2007} produces a non-normalised probability distribution for a parameter space by taking random steps through the space. The steps are randomly selected from a one-dimensional Gaussian in each parameter, with a variable step size, storing each model encountered on the way. A new model is accepted if its likelihood is greater than the previous model, while less likely models are accepted with a probability equal to the likelihood of the new model divided by the likelihood of the old one. If a new model is not accepted, the old model is repeated in the Markov chain. The choice of a Gaussian selection function is widespread, though not essential in MCMC; but it has the advantage that it favours small steps over large ones, and the probability of stepping between two points is identical in both directions. This second property ensures that the steps satisfy the Metropolis detailed balance condition~\citep{hastings1970,Metropolis1953}.

We use an MCMC method to explore models over a broad range of values for the halo parameters $\alpha$, $\beta$, $\gamma$, $r_{\rm{s}}$, and $v_{\rm{max}}$, and stellar disk parameters $f_{\Upsilon}$ and $f_{R_{\rm d}}$. A summary of the parameters and their ranges can be found in Table~ \ref{tblparams}. We use the October 2012 release of the publicly available CosmoMC code~\citep{lewis2002} as an MCMC driver. CosmoMC was originally designed to determine cosmological parameters from the cosmic microwave background, but it has a generic MCMC mode that we utilise here. We choose stepping mode 4 ``slow grid" - which is described in the CosmoMC documentation.

To facilitate the comparison between different parametric models, after the MCMC chains have been generated, we calculate the physical parameter $\gamma_{\rm in}$ - the logarithmic slope of the density profile at the location of the innermost radial bin of the measured rotation curve - as the basis for comparison. This allows us to explore the issue of whether our choice of a set of parametric profiles has a significant impact on the estimates of the log-slope of the halo profile. 

For each data set we generate 16 MCMC chains, each of which contains $\sim5\times10^5$ accepted models and has a different set of starting parameters. All results in this paper have more than $7\times10^6$ models. The multiple starting points help to ensure that a local maximum in the likelihood does not trap the algorithm - although due to the nature of MCMC this is unlikely to happen if the chains are run for sufficiently long. We are able to verify that MCMC chains have converged by comparing multiple chains (if converged they should all produce the same, smooth, distribution of values for each parameter), and all chains produced for this paper have been checked for convergence in this way. 

\subsection{Prior distributions and normalisation}
\begin{figure}
  \includegraphics[width=\linewidth]{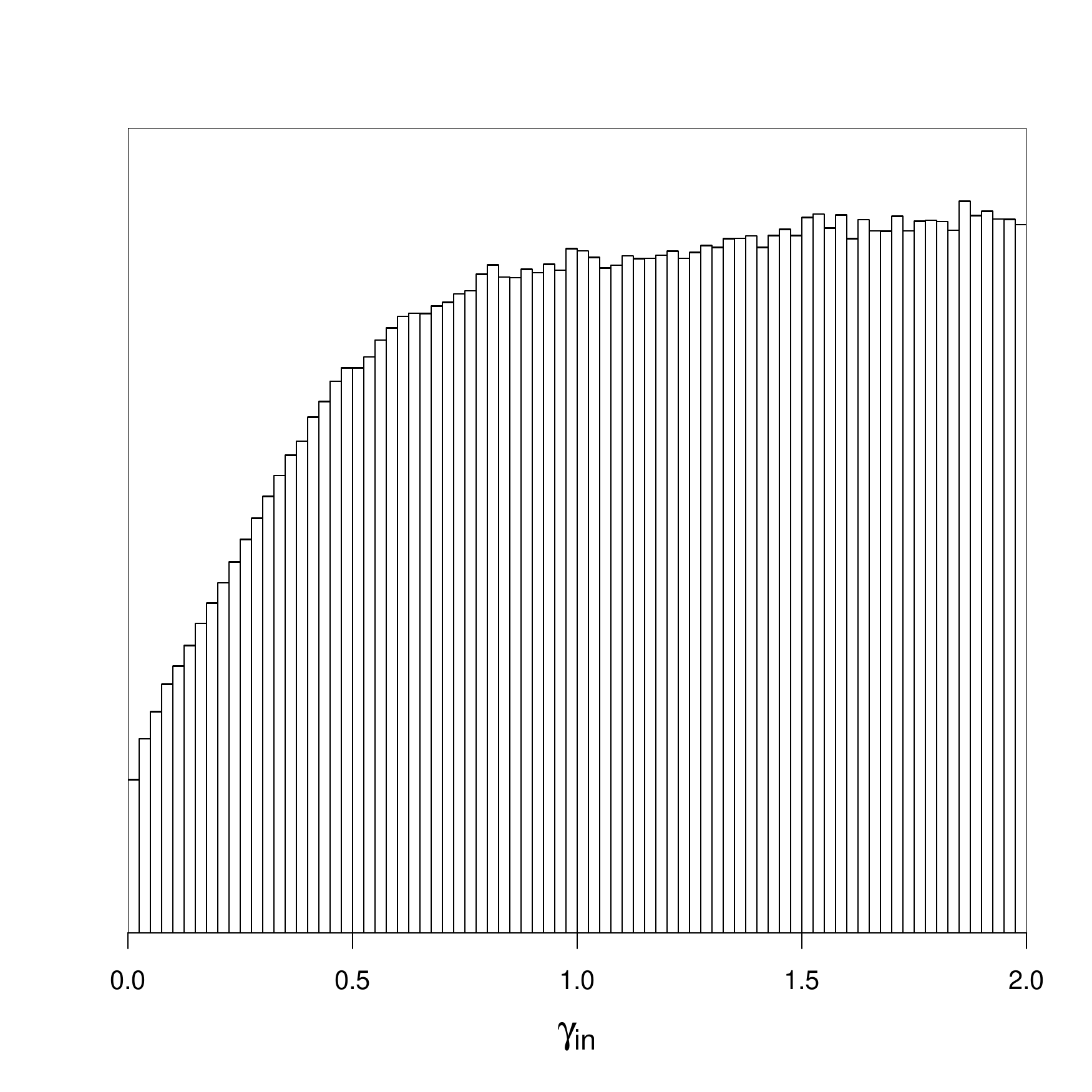}
  \caption{Histogram of $\gamma_{\rm in}$ for a ``flat" run in which all parameter combinations are equally probable, illustrating the need to normalise histograms for derived quantities.}
    \label{flatrun}
\end{figure}

We assume uniform priors on all our parameters and as a result the shape of our parameter space is simply a uniform hypercuboid. However the non-linear transformation of this space into physical quantities such as $\gamma_{\rm in}$ or ${\rm d}\log\rho(r) / {\rm d}\log r$ means that the prior distributions of these quantities may no longer be uniform. This can be seen from Fig.~\ref{flatrun} which shows the distribution of $\gamma_{\rm in}$ values generated by a ``flat" run that returns $\chi_{\rm red}^2=1$ for all models. There is a volume bias in ($\alpha,\beta,\gamma$) space away from $\gamma_{\rm in}=0$, due to the fact that the log slope only reaches zero at $r=0$ and any finite scale radius will shift $\gamma_{\rm in}$ away from zero even when $\gamma=0$. The derived parameter, $\gamma_{\rm in}$, therefore requires normalisation, because there is a one-to-many mapping between each value of $\gamma_{\rm in}$ and sets of parameter values, i.e. the parameter space for each value of $\gamma_{\rm in}$ is a different size. Without normalisation, this would bias a histogram of $\gamma_{\rm in}$ towards areas where the parameter space had a larger volume and away from $\gamma_{\rm in}=0$. In what follows, all histograms of derived quantities such as $\gamma_{\rm in}$ are normalised by the corresponding histogram obtained from a flat run. We have verified that this normalisation process is robust in the sense that normalising the output of one flat run by that of another leads to a uniform distribution to within the bin noise.

Although in some of the figures in later sections we plot certain parameters using logarithmic scales, we have chosen, after extensive testing, not to use a logarithmic range for any of our parameters. Typically, logarithmic spaces are used for parameters whose values can range over several orders of magnitude. However, for a parameter whose values vary over a limited range, a uniform prior in a logarithmic space corresponds to a prior in linear space that favours smaller values of the parameter. Since CosmoMC dynamically varies the step sizes for all parameters, it is already able to survey a large range in parameter space while concentrating on small values of particular parameters if necessary. We therefore use uniform priors in linear space for all our parameters, as this greatly simplifies the interpretation of the output from the MCMC chains.

The $\gamma=0$ boundary can lead to a bias of the MCMC chains away from models with very small values of $\gamma$. This effect came to light during early tests of our analysis and is alleviated by extending the range of $\gamma$ values to include $\gamma<0$ (the actual range used is $[-2,2]$) and using the modulus of $\gamma$ in calculations of the halo rotation curve. This ensures that the models of most interest to us are not in an unusual portion of the parameter space. Tests on synthetic data presented in Section~\ref{sec:tests} confirm that the algorithm is indeed able to recover halo profiles with $\gamma \sim 0$.

\section{Tests on synthetic data}
\label{sec:tests}

Before applying the algorithm to observed data, it is essential to demonstrate that it can successfully recover the properties of galaxies with known characteristics. To this end, we generate a number of synthetic data sets for each of two galaxy types: (1) a low surface brightness (LSB) galaxy, based on the galaxy DDO 154, for which we expect the algorithm to perform well; and (2) a high surface brightness (HSB) galaxy, based on NGC 7793, which we use to demonstrate explicitly the strong degeneracies which occur in modelling this class of galaxy and the way in which the MCMC approach naturally identifies their presence.

To generate the artificial data, we fit a rotation curve for a thin exponential disk~\citep{BT} to the stellar rotation curve provided by~\cite{deblok2008} (which uses a ${\rm sech}^2(z)$ vertical profile and is generated from surface brightness models using the ROTMOD task in Gypsy) , which includes estimates of the mass-to-light ratio. For DDO 154 and NGC 7793 we obtain disk scale lengths $R_{\rm d}$ of $2.15$ kpc and $2.51$ kpc, respectively. 

We next assume a particular dark matter halo profile for each galaxy, and calculate the corresponding circular speed curve which we add in quadrature to the stellar rotation curve, assuming the same stellar $\Upsilon_{\rm 3.6 \mu m}$ as \cite{deblok2008}. If we are allowing the stellar $\Upsilon$ to vary along the MCMC chains, the stellar circular speed at all radii is scaled by a factor $\sqrt{f_\Upsilon}$ before it is combined with the dark matter curve. Finally, the gas contribution is added (in quadrature) to obtain the total circular speed curve for our synthetic galaxy. The properties of the gas component are held fixed along the chains - we use the observed gas contributions in DDO 154 and NGC 7793 for the model LSB and HSB galaxies, respectively.

Before applying our algorithm to these synthetic data, we add Gaussian observational noise to each data point. The final data set passed to the algorithm is then the set of ``observed'' velocity data points and their error bars, and the contribution to the rotation curve of the stellar and gas disks, which are assumed to have been measured for the observed galaxy, in the manner which has been done by~\cite{deblok2008} for the THINGS sample.

\subsection {Reduced observational errors}
\label{testA}

\begin{figure}
  \includegraphics[width=\linewidth]{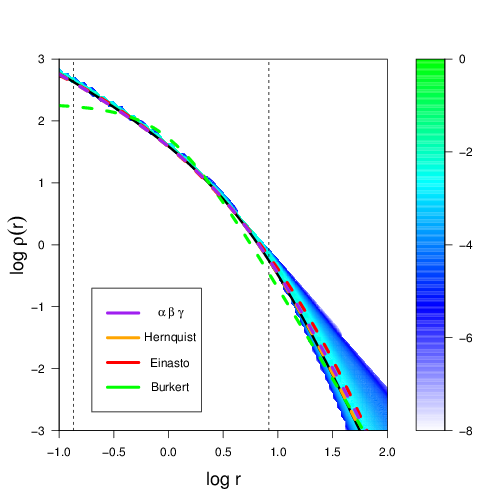}
  \includegraphics[width=\linewidth]{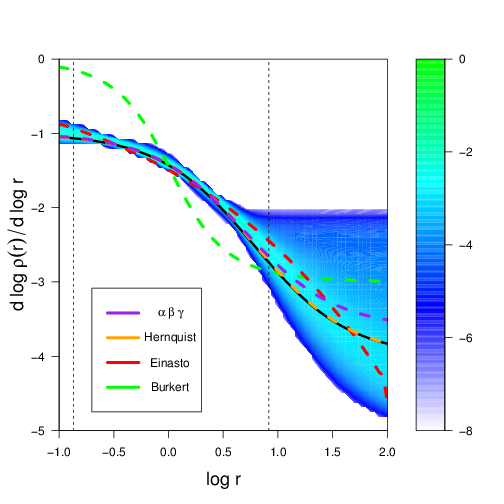}
    \caption{Results of the algorithm applied to artificial data for an LSB galaxy with small observational error bars. The colour scale is $\log [N(\mathbf{x}) / N_{\rm total}]$, i.e. the log proportion of models going through a particular point. \textbf{Top:} the distribution of halo density profiles for all models accepted in the MCMC chains. The vertical dashed lines show the inner and outer limits of the input data, and the coloured dashed curves show the best fits of single profiles:  purple for the ($\alpha,\beta,\gamma$) profile, green for the burkert profile, red for the Einasto profile, orange for the Hernquist profile. The black curve is the input profile. \textbf{Bottom:} the distribution of halo density profiles in (${\rm d}\log \rho(r) / {\rm d}\log r,\log r$) space i.e. the log slope with respect to log radius. Overlaid profiles follow the same colour scheme as in the top panel.}
    \label{fig:lowerrortest}
\end{figure}

In our first test, we explore the performance of the algorithm on almost ``ideal'' data by assuming fractional error bars of $0.02$ on the velocities. These are about a factor of five smaller than the typical errors on the THINGS rotation curves. The dark matter halo is assumed to be a Hernquist profile, i.e. $(\alpha,\beta,\gamma) = (1,4,1)$, with parameters given in Table~\ref{tab:testparams}. For this test, we consider an LSB galaxy whose stellar mass profile is based on the exponential disk fit to DDO 154.

Fig.~\ref{fig:lowerrortest} presents the results of this test. As the top panel shows, given very precise input data, the algorithm is able to recover very accurately the density profile of the dark matter halo. The lower panel shows the distribution of accepted models in the ${\rm d}\log \rho(r)/{\rm d}\log r$ versus $\log r$ plane, indicating that the log slope is very well modelled at all radii within the data range (indicated by vertical dashed lines in the plot.) The performance of the algorithm is facilitated by the low surface brightness of the stellar disk. Although in this instance we did not allow $f_{\Upsilon}$ to vary, the full range in values of this parameter  covers a range in amplitude of $\lesssim1$ \kms in the total rotation curve. It is important to note, however, that the distribution of models in Fig~\ref{fig:lowerrortest} marginalises over all the model parameters and shows that the algorithm takes advantage of the freedom provided by the many parameters to recover the overall density profile of the galaxy. Although the individual model parameters are not necessarily recovered correctly, this does not affect our conclusion that the algorithm is unbiased, as our goal is to recover physical quantities such as the log slope as a function of radius rather than values for parameters of our particular form for the halo profiles.

The results of this test show that the algorithm performs well in the idealised case of high quality data for an LSB galaxy. We find  no evidence of any modelling bias in either the shape or amplitude of the density profile and therefore conclude that the algorithm is working correctly. In the following sections, we consider its application to data sets of differing quality including realistic error bars that correspond to the quality of presently available rotation curves, as well as to data for HSB galaxies in which the uncertainty in the stellar mass to light ratio of the galaxy plays an important role.

\subsection {Realistic Observational Errors}
\label{testB}

\begin{figure}
  \includegraphics[width=\linewidth]{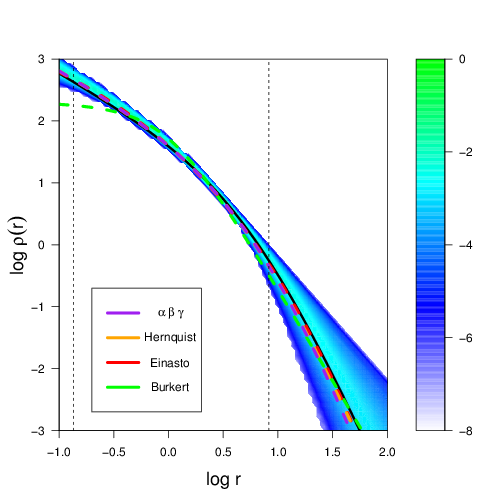}
  \includegraphics[width=\linewidth]{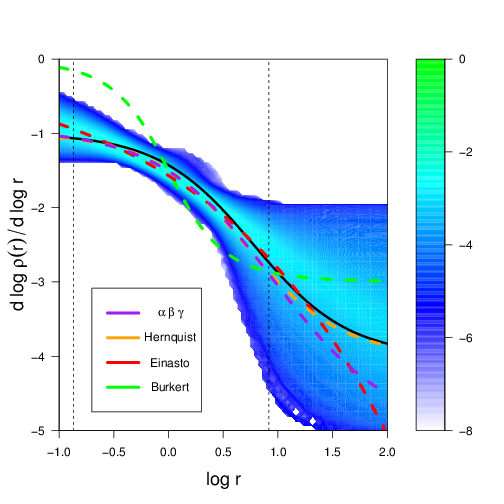}
    \caption{As in Fig.~\ref{fig:lowerrortest} but for artificial data for an LSB galaxy with realistic observational error bars. See text for a discussion.}
    \label{normalerrortest}
\end{figure}
\begin{figure}
  \includegraphics[width=\linewidth]{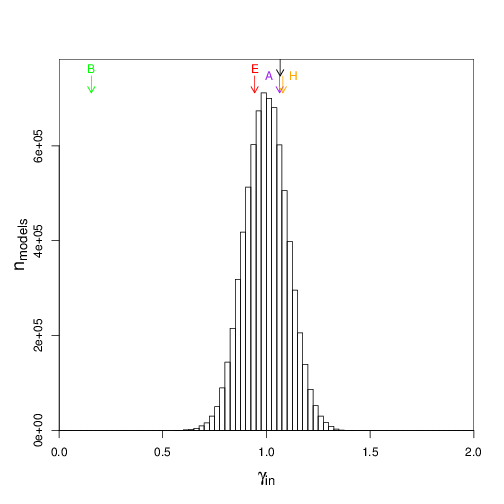}
  \includegraphics[width=\linewidth]{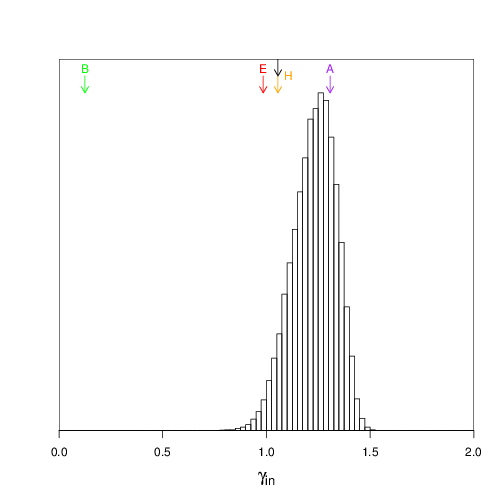}
    \caption{Histograms of $\gamma_{\rm in}$ for artificial data for an LSB galaxy (upper panel) and an HSB galaxy (lower panel). Arrows indicate the values of $\gamma_{\rm in}$ for the input model (black) as well as for the best-fit Burkert (green), Einasto (red), Hernquist (orange) and ($\alpha,\beta,\gamma$) (purple) profiles. See text for a discussion.}
    \label{HSBtest}
\end{figure}

\begin{figure}
  \includegraphics[width=\linewidth]{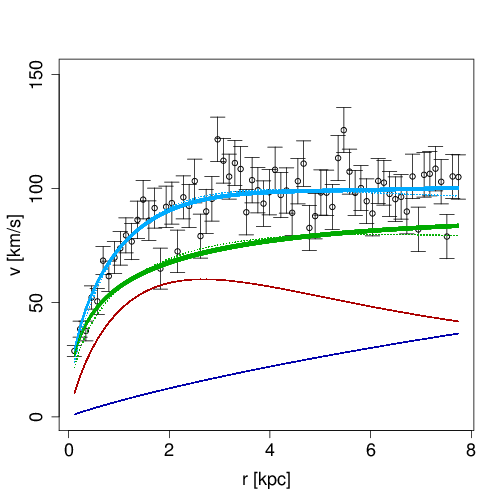}
    \caption{The highest likelihood model identified by the MCMC chains in the HSB galaxy test. The dotted curves are the input model, while the solid curves are for the models in the MCMC bin (in the 7-dimensional parameter space) containing most models (the thickness of the curve is due to many similar models being over-plotted). The stellar disk is shown in red, the gas disk is dark blue, the dark matter halo is green, and the expected (total) rotation curve is light blue.}
    \label{HSBtest2}
\end{figure}

In this test, the underlying galaxy model is identical to that used in the previous section, but we now add observational errors of $10$ percent to the total rotation curve which is more representative of the errors typical of current observed data sets, such as THINGS. Fig.~\ref{normalerrortest} presents the results of this test. As expected, the most obvious difference with the previous test is that the distribution of halo models is much broader. Nevertheless, as the top panel of the Figure shows, the algorithm accurately recovers the input halo profile over the radial range probed by the rotation curve data. Outside this region, the range of consistent models expands significantly. We emphasise that outside the range of the observed data, the apparent constraints on the halo profile are the result of our assumption of a parametric model for the dark matter halo. In the absence of physical constraints at larger or smaller radii than those probed by the data, we are effectively using all five halo parameters to fit the rotation curve over the finite radial range covered by the data. The physical relevance of our models is thus limited to the volume probed by the data and, as stated earlier, the values of the parameters are less significant than the overall profile obtained in the range of the data.

Of particular relevance to the question of resolving the cusp-core issue, the lower panel shows that we are still able to exclude models with uniform density cores at a high level of confidence even in the presence of realistic velocity errors. This is made more quantitative by the histogram in the upper panel of Fig.~\ref{HSBtest} which shows the distribution of the derived quantity $\gamma_{\rm in}$, the logarithmic slope of the halo density profile at the radius of the innermost data point. The various vertical arrows in the plot show the results of the direct fitting of a selection of individual halo models to the rotation curve data, rather than taking the MCMC approach. This is akin to the type of modelling that is more commonly applied to rotation curve data, and the models used span the range of models normal considered in such analyses. The peak of the MCMC histogram coincides with the values of $\gamma_{\rm in}$ for the directly fitted $(\alpha,\beta,\gamma)$, NFW and Einasto models. More importantly, the histogram shows that the input data (generated for a cusped halo model) are able to exclude Burkert halo profiles. This shows that our MCMC analysis is able to constrain the inner slope of the dark matter halo of an LSB galaxy.

The high surface brightness case shows a distribution that is more than $1\sigma$ from the input value, but as the best fit ($\alpha,\beta,\gamma)$ model still coincides with the peak of this distribution, this is not a problem with MCMC. Some realisations of the random errors may disfavour the input model compared to other ones, and the ``correct" distribution that the MCMC should output is that which is pointed to by the data, which does not always correspond to the input values because of this.

We conclude that current data for LSB galaxies are sufficient to place constraints on the dark matter profiles of those galaxies. We are able to make this stronger claim than previous work due to the greater generality of our modelling.

\subsection{High Surface Brightness galaxy}
\label{testC}

We now consider the case of an artificial data set for an HSB galaxy which we construct based on the higher surface brightness stellar contribution of NGC7793, but embedded in the same dark matter halo as in the previous sections so that the contribution to the simulated rotation curve from baryons is comparable to that of the dark matter. Because a maximal (or near maximal) disk could provide a reasonable fit to the data in this case (see Fig.~\ref{HSBtest2}), it is important to know whether or not it is still possible to constrain the dark matter halo parameters. 

As the bottom panel of Fig.~\ref{HSBtest} shows, the algorithm is able to obtain a constraint on the dark matter halo profile which appears to be only marginally weaker than that returned in the LSB test, as seen in the upper panel and in Fig.~\ref{normalerrortest}. However, the results for the HSB galaxy exhibit a bias towards steeper slopes for the reasons stated above. The MCMC distributions for $\gamma_{\rm in}$
can be well approximated by a Gaussian for both these test cases, and thus can be used to estimate the uncertainty on this value. We find $\gamma_{\rm in} = 1.00 \pm 0.11$ and $\gamma_{\rm in} = 1.25 \pm 0.11$, for the low and high surface brightness cases, respectively. For comparison, a simple maximum likelihood fitting of the ($\alpha,\beta,\gamma$) halo profile to these data yields $\gamma_{\rm in} = 1.06$ and $\gamma_{\rm in} = 1.31$, respectively, but without providing information about the distribution of allowed values. 

It is important to remember that in this test we have assumed that the $\Upsilon$ and $R_{\rm d}$ of the stellar components are both known and are therefore fixed at their correct values along the MCMC chains. Given the significant contribution of the baryons to the overall gravitational potential in this case, the results will clearly be strongly sensitive to uncertainties in these two parameters. In Section~\ref{testH} we return to this issue and consider the case in which the stellar parameters are only very weakly constrained.

\subsection {Higher resolution data sets}
\label{testD}

\begin{figure}
  \includegraphics[width=\linewidth]{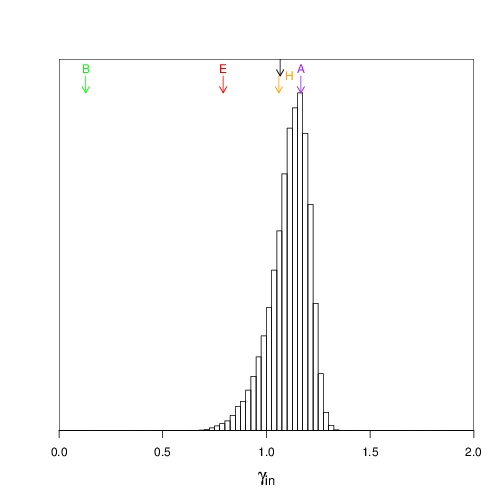}
    \caption{As in Fig.~\ref{HSBtest} but for artificial data for an LSB galaxy with realistic observational error bars, and with the radial sampling density of the rotation curve increased by a factor of two. See text for a discussion.}
    \label{highrestest}
\end{figure}

We now investigate what would happen if, in the future, data sets of higher spatial resolution became available. Thus far, our artificial data sets have used the same radial bins as those in the THINGS rotation curves from~\cite{deblok2008} for the two galaxies on which they are based (i.e. DDO 154 for the LSB case and NGC 7793 for the HSB case). In this test, we generate a data set with twice the number of radial bins over the same radial range. We assume that the velocity error bars remain unchanged - while this may be pessimistic (higher resolution sampling of the rotation curve might be expected to coincide with improved velocity resolution also), it allows us to separate the impact of higher spatial resolution rotation curves from more precise velocity measurements. The current radial bin size for DDO 154 is 136pc, which at a distance of 4.3Mpc is an angular resolution of $6.5''$ \citep{deblok2008}, so this higher density data set would have an angular resolution of $3.25''$.

Fig. \ref{highrestest} shows that a value of $\gamma_{\rm in} = 1.13 \pm 0.08$ is obtained from the high spatial resolution test, and whilst this gives a 27\% reduction in reported error, the peak of the distribution remains offset from the input model value of 1.07 by a similar amount to the lower spatial resolution test. This indicates that there is relatively little to be gained at the present time from increased spatial resolution until the velocity errors are significantly reduced.

\subsection {Data sets extending to larger radii}
\label{testE}

\begin{figure}
  \includegraphics[width=\linewidth]{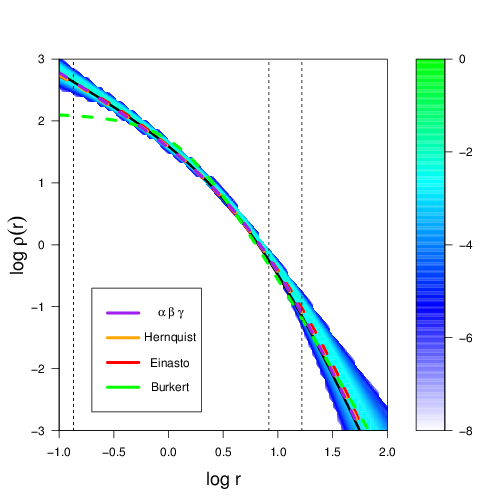}
  \includegraphics[width=\linewidth]{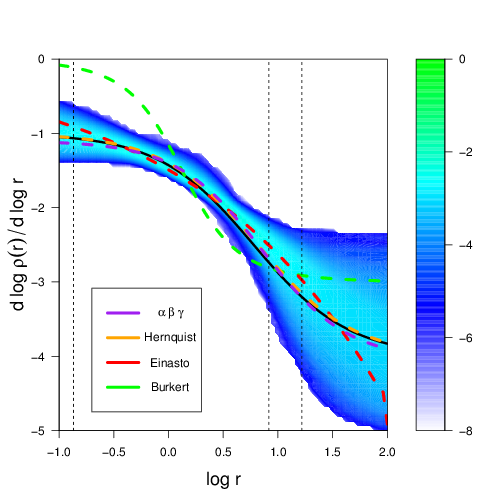}
    \caption{As in Fig.~\ref{fig:lowerrortest} but for artificial data for an LSB galaxy with realistic observational error bars and the data range extended by a factor of 2. The rightmost vertical dashed line represents the new maximum radius, centre vertical dashed line represents the original maximum radius in previous tests. See text for further discussion.}
    \label{extendedtest}
\end{figure}

\begin{figure}
  \includegraphics[width=\linewidth]{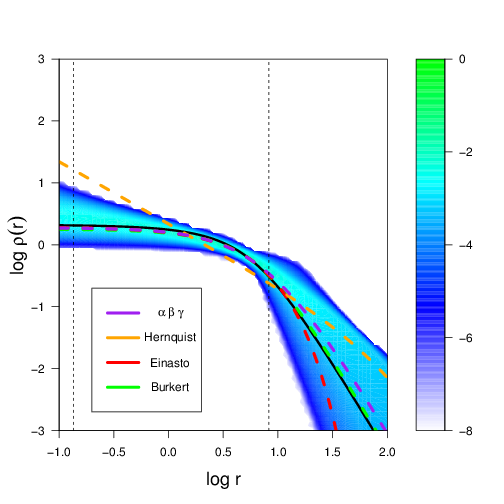}
  \includegraphics[width=\linewidth]{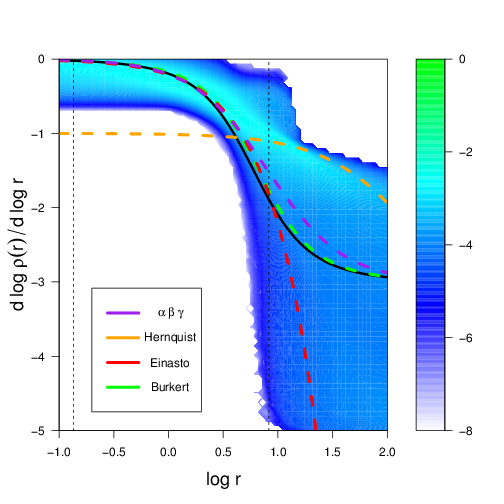}
    \caption{As in Fig.~\ref{fig:lowerrortest} but for artificial data for an LSB galaxy with a Burkert dark matter halo. }
    \label{coredtest}
\end{figure}
\begin{figure}
  \includegraphics[width=\linewidth]{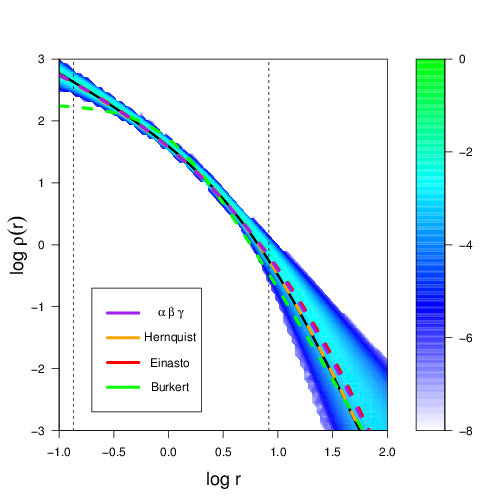}
  \includegraphics[width=\linewidth]{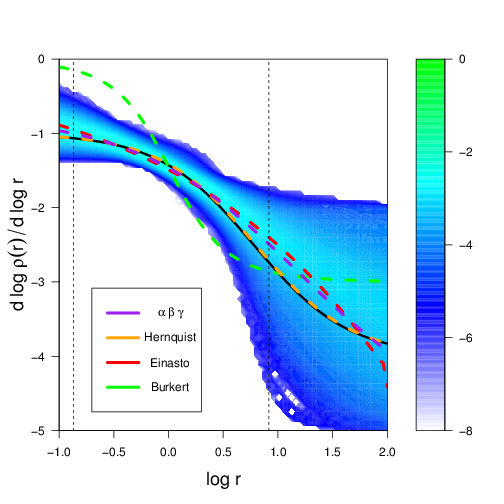}
    \caption{As in Fig.~\ref{fig:lowerrortest} but for artificial data for an LSB galaxy with realistic observational error bars, and free stellar disk parameters $f_{\Upsilon}$ and $f_{R_{\rm d}}$.}
    \label{Lowfreetest}
\end{figure}

\begin{figure}
  \includegraphics[width=\linewidth]{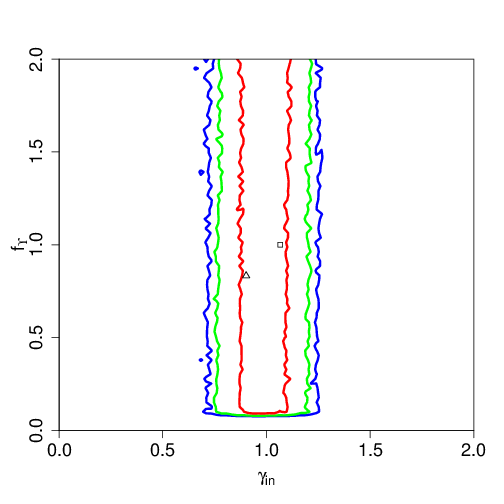}
    \caption{Contour plot of models for an artificial data set with a Hernquist input profile, free stellar parameters, and realistic errors, in the $f_{\Upsilon}$ versus $\gamma_{\rm in}$ plane. The red contour encloses the region containing $0.683$ of all models (equivalent to a $1\sigma$ contour), while the green and blue contours contain $0.95$ ($2\sigma$) and $0.999$ ($3\sigma$) of all models, respectively. The corresponding contours in the $f_{R_{\rm d}}$ versus $\gamma_{\rm in}$ plane show the same independence of $f_{R_{\rm d}}$ and $\gamma_{\rm in}$. The open triangle shows the model with the highest likelihood in the MCMC output, while the open square shows the input model.}
    \label{gammainml}
\end{figure}

\begin{figure}
  \includegraphics[width=\linewidth]{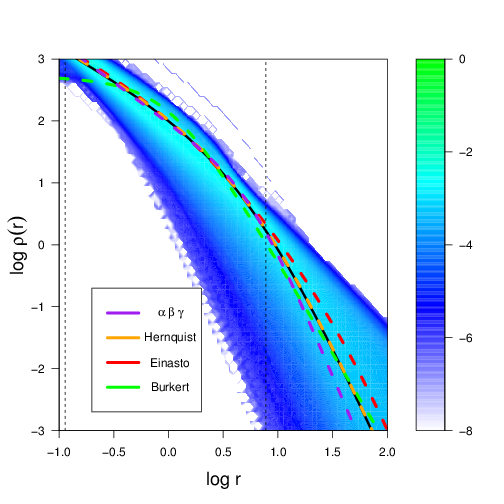}
  \includegraphics[width=\linewidth]{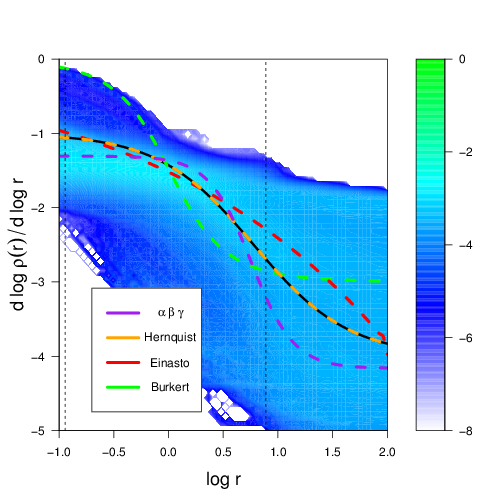}
    \caption{As in Fig.~\ref{fig:lowerrortest} but for artificial data for an HSB galaxy with realistic observational error bars, and free stellar disk parameters $f_{\Upsilon}$ and $f_{R_{\rm d}}$.}
    \label{Highfreetest}
\end{figure}

In this section, we apply our algorithm to data sets which extend to significantly larger radii than are probed by current rotation curve observations. We implicitly assume that the target galaxy has sufficient gas at these radii to make the observations possible, and that the baryonic disk remains relatively undisturbed so that our modelling assumptions remain valid. In principle, given a suitable target and sufficient observing time, it would be possible to obtain such a data set using existing instruments. 

The top panel of Fig.~\ref{extendedtest} shows the constraints on the density profile which we obtain for a data set extending to ~16kpc with velocity errors typical of current observations. Comparing this with Fig.~\ref{normalerrortest}, one might be tempted to conclude that improved constraints are obtained at all radii, including the inner regions. It is important to note that this occurs only because the data at large radii place tighter constraints on parameters in our halo model such as $\beta$ which leads to a reduced range of available parameter space at all radii. A non-parametric approach to the modelling would alleviate this situation, and we intend to pursue this in the future.

Comparison of Figs.~\ref{extendedtest} and~\ref{fig:lowerrortest} shows that the fit quality is not improved by as much as it is when the velocity errors are reduced. We conclude that  future observational programs should concentrate on reducing the velocity errors rather than increasing the spatial resolution or radial extent of the rotation curve data.

\subsection{Cored Profile Test}
\label{testF}

Much of the original impetus for detailed rotation curve modelling came from interest in the question of whether galaxy haloes are cored or cusped. In the previous sections, our input data were generated from model galaxies with Hernquist dark matter haloes, which are cusped ($\gamma = 1$). We now consider the case of a galaxy whose halo has a Burkert profile, a widely-used cored halo profile, which also forms the basis of the Universal Rotation Curve analyses of galaxy rotation curves~\citep[e.g.][]{ps1988, ps1991,pss1996}. Another reason for selecting this particular cored profile is that it is not a member of the ($\alpha,\beta,\gamma$) family (although over a restricted range of radii it can be closely represented by a profile from this set) and thus constitutes a test of the performance of the algorithm when the parameterisation of our models does not match the actual form of  the dark matter halo. We assume radial sampling similar to that of DDO 154 and realistic velocity errors.

The results of this test are shown in Fig.~\ref{coredtest}. The distributions in both panels show that the MCMC chains are favouring models with shallow inner profiles. Comparison with the results in section \ref{testB} shows that the algorithm is able to distinguish between haloes with cored and cusped density profiles in the radial range probed by the rotation curve. More quantitatively, the MCMC chains yield an estimate of $\gamma_{\rm in} = 0.15 \pm 0.13$, which excludes haloes with cusp slopes steeper than $0.67$ with $4\sigma$ confidence.

We conclude that our algorithm is able to distinguish between cored and cusped halo profiles using extant observational data sets for LSB galaxies.

\subsection{Free stellar disk parameters}
\label{testG}

We explore the impact of observational and modelling uncertainties in the $\Upsilon$ of the stellar population and the disk scale radius $R_d$ by repeating our analysis from section \ref{testB} but now allowing both quantities to vary along the chains. This represents the fact that the stellar population of the target galaxies is not known precisely, which translates into an uncertainty in the disk $\Upsilon$, and the simplification we make in the disk modelling by fitting a smooth function to the observed data, which may lead to uncertainty in the disk scale length $R_{\rm d}$.

We note that the distribution of $f_{\Upsilon}$ values from the MCMC chains is almost uniform, which indicates that the data are not able to constrain the $f_{\Upsilon}$ in this case. Nevertheless, Fig.~\ref{Lowfreetest} shows that the algorithm is able to constrain the halo density profile with a similar level of precision to that obtained when the stellar parameters were held fixed (Fig.~\ref{normalerrortest}). This behaviour can be understood by the fact noted earlier that the stellar disk contributes to the total rotation curve in quadrature, rather than linearly. As a result,  the full range in disk $f_{\Upsilon}$  covers a range in amplitude of $\lesssim1$ \kms in the total rotation curve.

Our test shows that a galaxy with a comparable ratio between surface brightness, observed rotation curve, and observation errors to the one we have synthesised here cannot constrain $f_{\Upsilon}$, but can still constrain properties of the dark matter halo. 

Fig.~\ref{gammainml} presents contours in the $f_{\Upsilon}$ versus $\gamma_{\rm in}$ plane and shows explicitly that the estimate of $\gamma_{\rm in}$ is independent of $f_{\Upsilon}$. The corresponding contours for $f_{R_{\rm d}}$ are very similar to those for $f_{\Upsilon}$. This emphasises the power of the MCMC approach over simple model fitting. If the MCMC analysis of a real galaxy results in a flat distribution of the disk parameters whilst still being able to constrain dark matter parameters, and exhibiting no correlation between $\gamma_{\rm in}$ and the disk parameters, we are able to conclude with confidence that the halo parameters have genuinely been constrained independently of the disk.

\subsection{High surface brightness test with free stellar parameters}
\label{testH}

Distinct from the question of whether a near maximal disk makes it impossible to constrain the properties of a dark matter halo, we wished to investigate the degeneracy between $v_{\rm max}$ and $f_{\Upsilon}$, both of which affect the amplitude of the total rotation curve. We therefore applied the algorithm with free $f_{\Upsilon}$ and $f_{R_{\rm d}}$ to our HSB data set. The results are shown in Fig.~\ref{Highfreetest}. As one would naively expect, the degeneracy between  $v_{\rm max}$ and $f_{\Upsilon}$ means that a much wider range of halo profiles may be consistent with the observed data set. 

We therefore conclude that HSB galaxy rotation curves are of limited value for constraining dark matter halo profiles unless robust information about the stellar disk mass distribution is available either through stellar population-based estimates of $\Upsilon$ or estimates of the vertical velocity dispersion within the disk~\citep[see e.g.][]{Bershady2010}. This agrees with previous work: for example \cite{gentile2004} who found that Burkert and NFW haloes yielded equally good fits to the data for NGC 7339 which they partly attributed to uncertainties in $\Upsilon$ and partly to the limited radial extent of the observed data.

\begin{figure}
 \includegraphics[width=\linewidth]{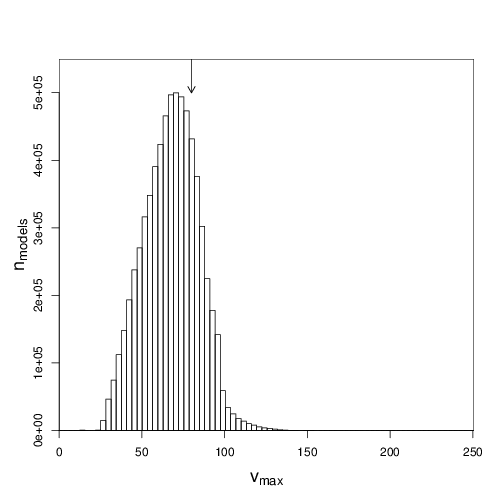}
 \includegraphics[width=\linewidth]{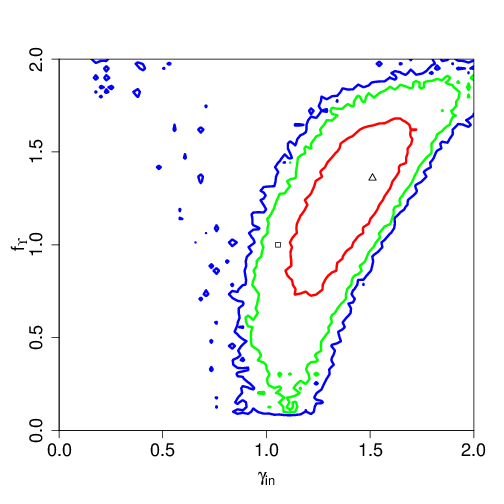}
 \caption{\textbf{Top:} The constraint of $v_{\rm max}$ in the HSB test. \textbf{Bottom:} Contour plot showing the correlation between $f_{\Upsilon}$ and $\gamma_{\rm in}$ for the same model. Contours and symbols are as in Fig.~\ref{gammainml}.}
 \label{hernHSBcorr}
\end{figure}

Compared to the test in Section~\ref{testC}, Fig.~\ref{hernHSBcorr} shows that we obtain a much weaker constraint on $v_{\rm max}$ here, which is an indication that the dark matter component is highly degenerate with the stellar component, and thus the result for the shape of the dark matter halo is unreliable. The lower panel of Fig.~\ref{hernHSBcorr}, presents a contour plot showing the degeneracy between $f_{\Upsilon}$ and $\gamma_{\rm in}$ for this test. The input value $\gamma_{\rm in} = 1.05$ is much more strongly excluded for $f_\Upsilon=1.5$ than for $f_\Upsilon=1.0$. If such a degeneracy is seen when working with real data, it must be resolved before the likelihood distribution produced can be accepted. 

In this case, we obtain an estimate of $\gamma_{\rm in} = 1.36 \pm 0.19$, giving the impression that the data are able to distinguish strongly cusped haloes from uniform cores. It is worth noting that the MCMC peak is consistent with the best fit $(\alpha, \beta, \gamma)$ curve - both exclude the input value of $\gamma_{\rm in}$ at $\approx 2\sigma$. The power of our MCMC approach to the problem is that we are alerted to the fact that these constraints may be spurious by the presence of strong degeneracies between parameters. We can thus explicitly and straightforwardly identify those cases in which the algorithm fails to constrain the halo parameters of the target galaxy. A simple fitting of a small set of halo profiles, on the other hand, could yield artificial constraints on the halo profile which were not actually arising from the properties of the data themselves. 

\subsection{Over-estimated errors}
\label{sec:correl}
\begin{figure}
 \includegraphics[width=\linewidth]{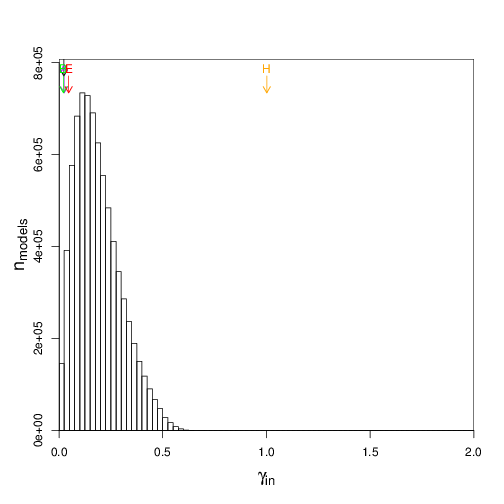}
 \includegraphics[width=\linewidth]{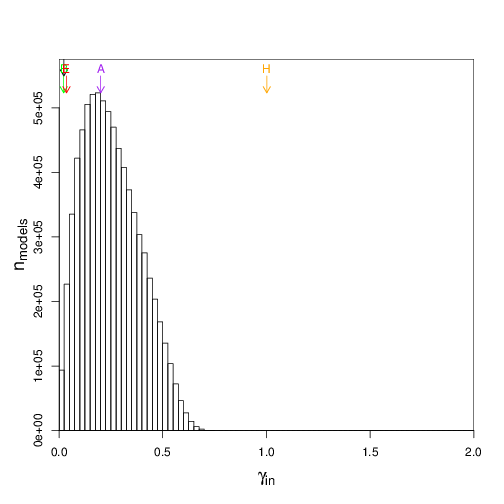}
 \caption{\textbf{Top:} Histogram of $\gamma_{\rm in}$ obtained for a data set generated for a galaxy with a Burkert model halo and realistic observational errors (Model F). \textbf{Bottom:} As in top panel, but with the error bars inflated by a factor of four relative to the actual Gaussian noise added to the data. Arrows are as in Fig.~\ref{HSBtest}.}
 \label{burkcorr}
\end{figure}

\begin{figure}
 \includegraphics[width=\linewidth]{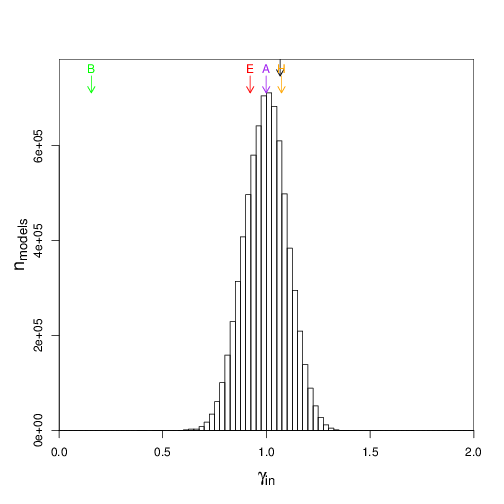}
 \includegraphics[width=\linewidth]{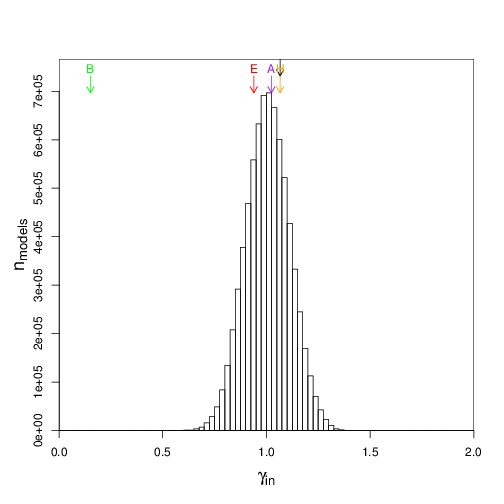}
 \caption{As in Fig.~\ref{burkcorr} but for data generated for a galaxy with a Hernquist dark matter halo.}
 \label{herncorr}
\end{figure}

In our final test, we explore the performance of our algorithm on data sets in which the error bars do not reflect the scatter between successive data points along the rotation curve. This is seen in several of the THINGS galaxies, and is caused by the definition of the error bars in terms of the variation between the rotation curve on either side of the rotation axis rather than purely the statistical noise around an annulus of the tilted ring model. Systematic differences between the rotation curve on either side of the axis can therefore lead to over-sized errors being quoted for the data, resulting in a series of data points whose error bars are much larger than the scatter between the points.

To test the effect of such error inflation, we generated Hernquist-based and Burkert-based models where the error bars on the data points were a factor of four larger than the noise which had been added to the observation - the noise added was the same as that added in Model B. All other model inputs remain the same. The noise added to the data is generated randomly for each realisation, rather than being the same errors scaled differently for each model.

Figure \ref{burkcorr} shows, for the Burkert models, that the distribution of $\gamma_{\rm in}$ values returned by the MCMC chains is not greatly affected even by this significant level of over-estimation - the peak of the distribution moves by $\lesssim 0.1$ and the shape of the distribution is somewhat broadened. In the case of a galaxy with a Hernquist halo, Figure \ref{herncorr} shows that the effect on the distribution is even smaller. It is worth noting that in case of the Burkert halo, the best single fit of an ($\alpha,\beta,\gamma$) halo (indicated by the purple arrow) moves considerably, whilst the peak of the distribution from the MCMC algorithm does not. This is a further demonstration of the robustness of our MCMC method.

For the Hernquist case, the correct error case gives a log slope $\gamma_{\rm in}=1.00\pm0.11$ and the over-estimated error case gives a log slope $\gamma_{\rm in}=1.01\pm0.11$. In the Burkert case, the log slope changes from $\gamma_{\rm in}=0.15\pm0.11$ to $\gamma_{\rm in}=0.22\pm0.16$, which is clearly within the $1\sigma$ uncertainty.

\section{Comparison with previous methods}
\label{sec:comparison}

\subsection{Profile fitting}
\label{proffit}

Before we apply our algorithm to the observed data for DDO 154, we carry out a quantitative comparison between performance of our algorithm and that of the standard model fitting approach most common in the literature. In these tests, we fit five individual halo profiles which are commonly used in previous work. For each profile we used the {\tt optim} function in R\footnote{http://www.r-project.org/} which employs a limited-memory BFGS method with box constraints~\citep[L-BFGS-B; ][]{rmethod}. This method constrains the fitting to a specific parameter volume which we set to be the same parameter volume as that available to the MCMC algorithm in order to facilitate a fair comparison between the two approaches. We compare the best-fit model parameters returned by this method with (i) The model from the centre of the most favoured bin of the MCMC distribution (with 512 bins per parameter); (ii) the distribution of models returned by the MCMC analysis. 

In these comparison tests, we use data generated using two input halo profiles: the Hernquist profile from the previous sections, as an example of a cusped halo, and the Burkert profile as a representative cored halo. For consistency with previous work which typically holds the $\Upsilon$ and $R_d$ parameters constant, in this section we do not allow the stellar properties to vary along our MCMC chains. 

\begin{figure}
  \includegraphics[width=\linewidth]{figuresModelBgammainhist.\imgext}
    \includegraphics[width=\linewidth]{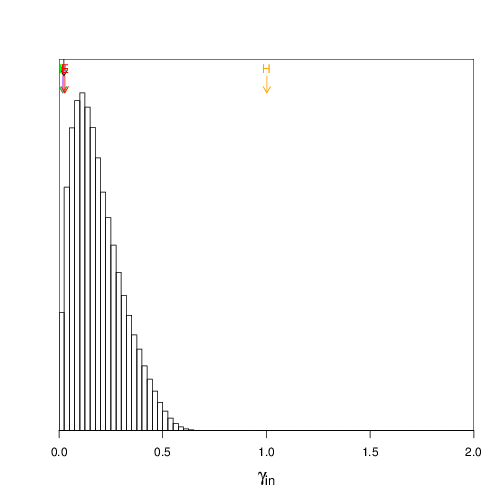}
  \caption{Histogram of $\gamma_{\rm in}$ values for a test with a Hernquist profile input (top) and a Burkert profile (bottom). Arrows indicate the values of $\gamma_{\rm in}$ for the input model (black) as well as for the best-fit Burkert (green), Einasto (red), Hernquist (orange) and ($\alpha,\beta,\gamma$) (purple) profiles. See text for a discussion.}
    \label{gammazero}
\end{figure}

The top panel of Fig.~\ref{gammazero} compares the histogram of $\gamma_{\rm in}$ values returned by the MCMC analysis with the values obtained from the fitting of individual halo profiles. The consistency of the $\gamma_{\rm in}$ value for the best-fit ($\alpha,\beta,\gamma$) model (purple arrow) with the peak of the MCMC-generated histogram, combined with an inspection of the MCMC chains for convergence, shows that the MCMC works with this data. The Hernquist, NFW and Einasto profiles yield also fits which are close to the MCMC peak.

More importantly, in this test the Burkert profile returns a $\gamma_{\rm in}$ value which is disfavoured by the MCMC distribution at the $3\sigma$ level. Reassuringly, its $\chi_{\rm red}^2$ values also disfavour these models relative to the NFW fit at the $3\sigma$ confidence level. This test suggests that the widely-used, simple curve-fitting approach can yield reliable results for cusped profiles, both in terms of recovering physical parameters and distinguishing between different halo models.  

The $\chi_{\rm red}^2$ values for all single fits, and the parameters for these fits and for the optimal MCMC result, are shown in Table \ref{tblfits}. In both cases the fits for the best MCMC result and the best-fitting ($\alpha,\beta,\gamma$) model find very similar values for $\gamma_{\rm in}$, even though there is variation in the parameters. 

\begin{figure}
  \includegraphics[width=\linewidth]{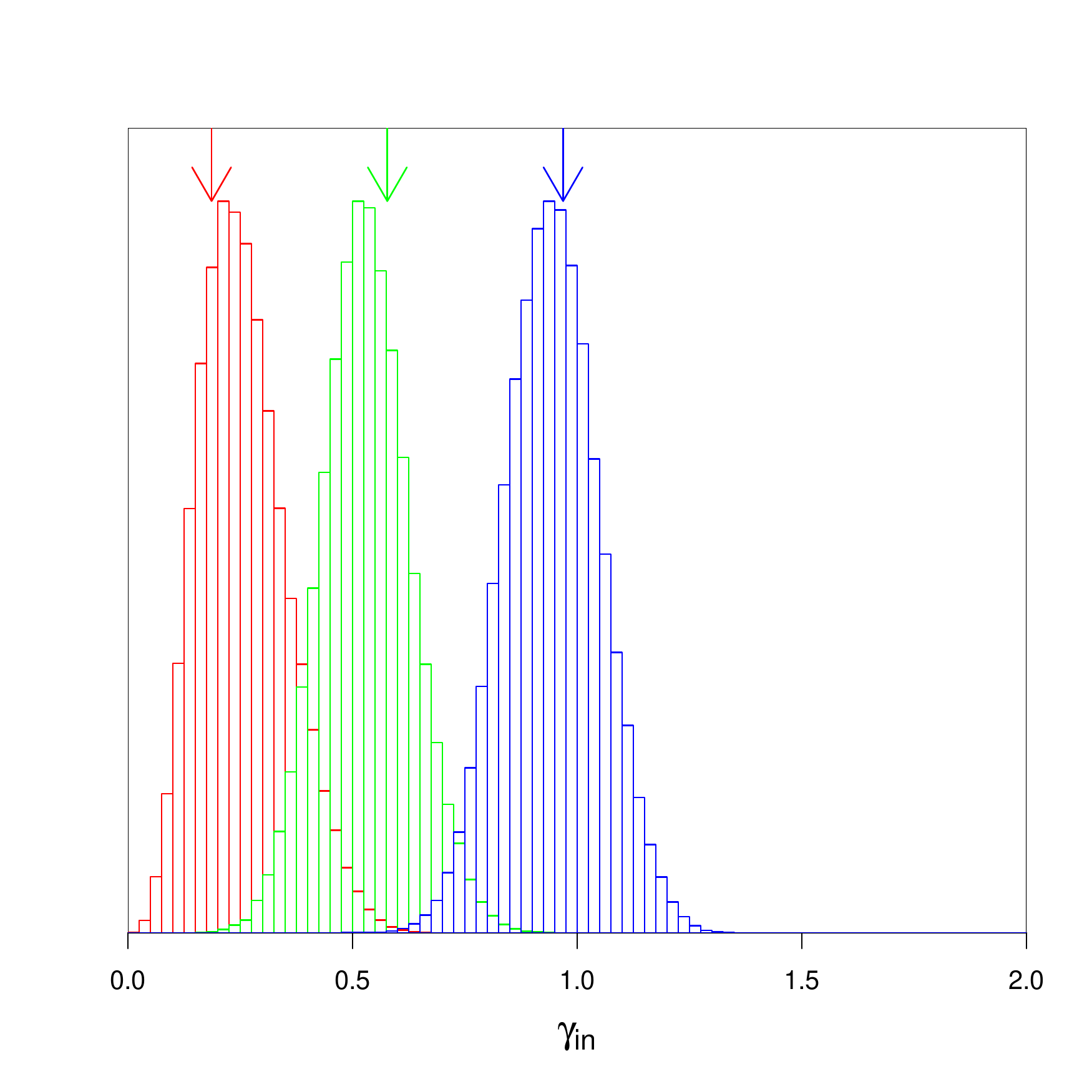}
    \caption{The constraint on $\gamma_{\rm in}$ for input values of $\gamma =(0.1,0.5, 0.9)$ which are shown in red, green and blue respectively. The arrows show the values of $\gamma_{\rm in}$ for the input model - note that these are not the same as the input values of $\gamma$.}
    \label{abgtest}
\end{figure}

In order to test the ability of our algorithm to distinguish between models with different values of $\gamma_{\rm in}$ we ran a series of models, based on Hernquist profiles $(\alpha, \beta, \gamma) = (1, 4, 1)$ but with $\gamma$ values ranging from 0.1 to 0.9 in steps of 0.1 - uniform cored and $\gamma=1$ cusped models having already been tested above. The distributions of $\gamma_{\rm in}$ for the three cases $0.1, 0.5$ and $0.9$ are presented in Fig. \ref{abgtest}. As the Figure shows, in each case, the MCMC chains are able to exclude both other values of $\gamma_{\rm in}$ at approximately $2\sigma$ confidence. We therefore conclude that provided the MCMC chains are properly converged, current observational data are sufficient to constrain $\gamma_{\rm in}$ to within $\lesssim \pm 0.25$. In addition to providing estimates of particular parameters, however, the MCMC approach returns the distributions of models which are consistent with the data, and is thus more informative than pure curve-fitting approaches.

\begin{table*}
\begin{tabular}{ | c | c | c | c | c | c | c | c | c | c | c |}
  Input Profile & Fitted Profile & $\chi^2_{\rm red}$ & $r_s$ & $\rho_s$  & $v_{\rm max}$ & $n$ & $\alpha$ & $\beta$ & $\gamma$ & $\gamma_{\rm in}$ \\   
  &&& (kpc) & ($10^{-3}$M$_{\odot}$ pc$^{-3}$) & (km/s) &&&&& \\
\hline  
ModelB & ($\alpha,\beta,\gamma$) (via MCMC) & $-$ & 3.84 & 102.485 & 48.085 & $-$ & 1.86 & 3.409 & 0.73 & 1.105 \\
ModelB & $(\alpha,\beta,\gamma$) (via fitting) & 1.325 & 9.09 & 10.425 & 48.141 & $-$ & 1.216 & 5 & 0.94 & 1.064 \\
ModelB & Burkert & 2.091 & 1 & 206.215 & 48.969 & $-$ & $-$ & $-$ & $-$ & 0.156 \\
ModelB & Einasto & 1.276 & 2.597 & 7.308 & 47.945 & 3.927 & $-$ & $-$ & $-$ & 0.943 \\
ModelB & Hernquist & 1.26 & 5.024 & 13.59 & 48.136 & $-$ & 1 & 4 & 1 & 1.079 \\
\hline
ModelF & ($\alpha,\beta,\gamma$) (via MCMC) & $-$ & 10.841 & 2.583 & 41.962 & $-$ & 1.157 & 2.894 & 0.004 & 0.067 \\
ModelF & $(\alpha,\beta,\gamma$) (via fitting) & 0.967 & 8.061 & 1.876 & 33.947 & $-$ & 0.799 & 3 & 0 & 0.018 \\
ModelF & Burkert & 0.918 & 6.667 & 1.807 & 30.558 & $-$ & $-$ & $-$ & $-$ & 0.021 \\
ModelF & Einasto & 0.932 & 9.063 & 0.271 & 29.339 & 0.982 & $-$ & $-$ & $-$ & 0.028 \\
ModelF & Hernquist & 2.55 & 220.171 & 0.01 & 57.227 & $-$ & 1 & 4 & 1 & 1.002 \\
\hline
\end{tabular}
\caption{Comparison of results obtained by fitting individual halo profiles to the data for Model B and Model F with the best-fit bin returned by the MCMC chains.}
\label{tblfits}
\end{table*}

\subsection{MCMC approaches}
The application of MCMC techniques to the analysis of rotation curves was previously carried out by~\cite{puglielli2010}. Their analysis was of a single galaxy, NGC 6503, using a more complex stellar model than is considered here (and, as a consequence, a higher dimensional parameter space). They used 18 parameters, and tested 4 distinct scenarios which effectively expanded the volume further, but without continuously parameterising the extra dimensions. The majority of the parameters related to the stellar distribution, while the dark matter halo was characterised by just three parameters: a density scale, a radial scale and an inner logarithmic slope. The other shape parameters of the halo were held fixed ($\alpha=1$ and $\beta=3$). They rejected the use of HI rotation curve data on the basis that the one dimensional data they were modelling could not differentiate between circular and non-circular motion. They instead use stellar rotation curves, which require additional calculation to take into account asymmetric drift. 

Our motivation for allowing the parameters $\alpha$ and $\beta$ to vary, in contrast to the approach of~\cite{puglielli2010}, is that we wish to determine explicitly the extent to which the observed data can constrain the halo profiles. By using an MCMC approach, parameters which are unconstrained are easily identified as such. Further, MCMC permits easy identification of parameter degeneracies. This is a key advantage of the method as it makes it possible to determine which physical quantities are constrained by the data. It is important to note that although individual parameters in our models (e.g. the asymptotic log slope of the halo profile $\gamma$) may not be constrained, the data may nevertheless constrain physical quantities such as the log slope at the innermost data point (which is a non-linear function of all the model parameters). A simple example of this is that the gravitating mass within the outermost data point, which depends on all five model parameters, can be constrained simply by virtue of our assumption that the gas is moving on circular orbits at each radius. If we fix the shape of the halo at large radii, however, we limit the ability of the models to reproduce the data and may therefore bias the estimate of the mass. Including more parameters is justified provided that the MCMC results are interpreted with caution - no credence is given to individual parameters which are unconstrained, and all parameter correlations are carefully explored.

With the availability of two dimensional velocity fields, non-circular motion can now either be removed from HI data, or target galaxies can be selected on the basis of their exhibiting low levels of non-circular velocity. In either case, gas is a more reliable source for rotation curves as it does not require further modelling. This partly motivates our use of simpler models of the stellar components in this work than those used by~\cite{puglielli2010}. Additionally, however, we have found from our study that the inclusion of parameters which relate to physical quantities that are not constrained by the observed data creates the risk of biasing the analysis due to complications in the shape of the parameter space or the presence of degeneracies between these parameters and physical quantities which the data can constrain. In particular, if there is a region of the parameter space where one or more of the parameters no longer has an influence on the model likelihoods, and models in this region have an adequate (but not optimal) likelihood, then the overall distribution of models may be biased towards this region and away from higher likelihood models. For example, if a maximal disk is consistent with the data then a model which has a low value of $v_{\rm max}$ can have a wide range of $(\alpha,\beta,\gamma,r_{\rm s})$ without affecting its likelihood. The volume of parameter space associated with such high-$\Upsilon$, low-$v_{\rm max}$ models is therefore increased, which may lead the MCMC chains to become biased in favour of maximal disk models.

We have therefore elected to take a Òbottom-upÓ approach to the stellar modelling rather than an Òall-inÓ approach and our models contain a minimal set of parameters. For the stellar distribution, two parameters is the least that are required to reflect the uncertainties in the data, and for the dark matter halo the $(\alpha,\beta,\gamma,v_{\rm max}, r_{\rm s})$ parameterisation is necessary to ensure that the full range of models which have previously been considered can be approximated within a single family.

\section{Application to DDO 154}
\label{sec:ddo154}

Having shown that our algorithm is able to produce reliable results, we now present an application to observational data for the LSB galaxy DDO 154. DDO 154 is a dwarf irregular galaxy whose mass has been estimated to be $3.0 \times 10^9$ M$_\odot$ inside a radius of 8kpc \citep{carignan1998}. The galaxy also has an extensive gas disk, providing good data for the study of its rotation curve. In a recent study of the rotation curve of DDO 154, also based on THINGS data, \cite{oh2011} found the log slope of its halo based on the inner data points to be $\alpha=0.29\pm0.15$. This study excluded the innermost data point, where our $\gamma_{\rm{in}}$ is defined, but in what follows we take this as a general description of the inner slope for comparison with our results.

It is worth noting that the error bars on the rotation curve of DDO 154 presented in \cite{deblok2008} appear, on inspection, larger than the noise in the data points, suggesting that they are not purely statistical errors and may be affected by the error inflation discussed earlier, in Section~\ref{sec:correl}. While we have shown already that such outsized errors do not strongly impact on the performance of our algorithm for an LSB galaxy, it is an undesirable feature of the tilted ring modelling approach to the determination of rotation curves which we plan to address in future work.

\begin{figure*}
  \includegraphics[width=\linewidth]{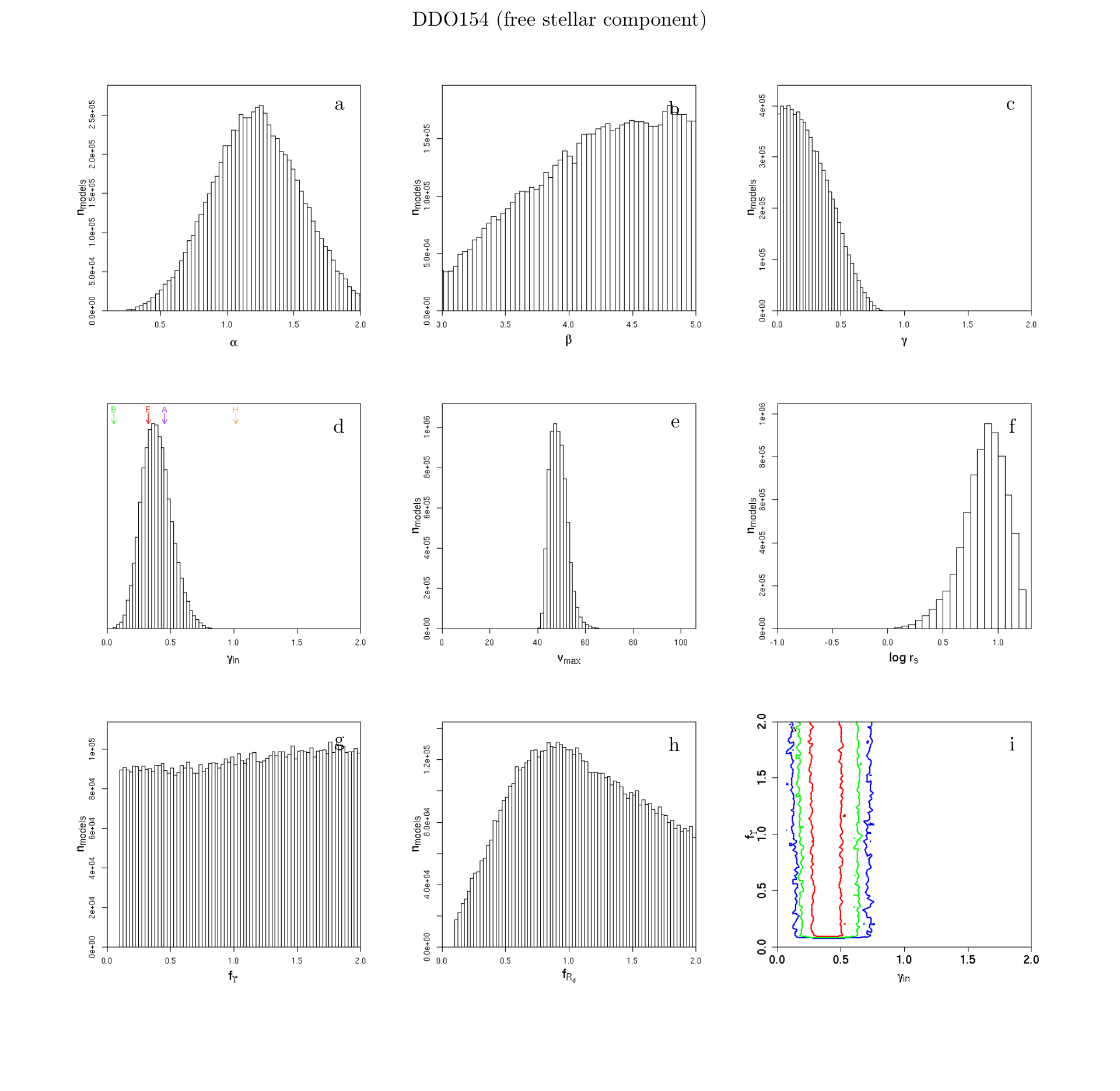}
  \caption{Detail of the output of the MCMC algorithm run on DDO 154 with free stellar disk parameters $f_{\Upsilon}$ and $f_{R_{\rm d}}$. \textbf{a)} Histogram of $\alpha$ values. \textbf{b)} Histogram of $\beta$ values.  \textbf{c)} Histogram of $\gamma$ values. \textbf{d)} Histogram of $\gamma_{\rm in}$ values.   Arrows indicate the values of $\gamma_{\rm in}$ for the best-fit Burkert (green), Einasto (red), Hernquist (orange) and ($\alpha,\beta,\gamma$) (purple) profiles.  \textbf{e)} Histogram of $v_{\rm max}$ values.  \textbf{f)} Histogram of $\log r_{\rm s}$ values.  \textbf{g)} Histogram of $f_{\Upsilon}$ values.  \textbf{h)} Histogram of $f_{R_{\rm d}}$ values. \textbf{i)} As in Fig.~\ref{gammainml} but for the case of DDO 154, showing that the constraint on $\gamma_{\rm in}$ is independent of $f_\Upsilon$.}
  \label{paramsddo154free}
\end{figure*}

\begin{figure}
  \includegraphics[width=\linewidth]{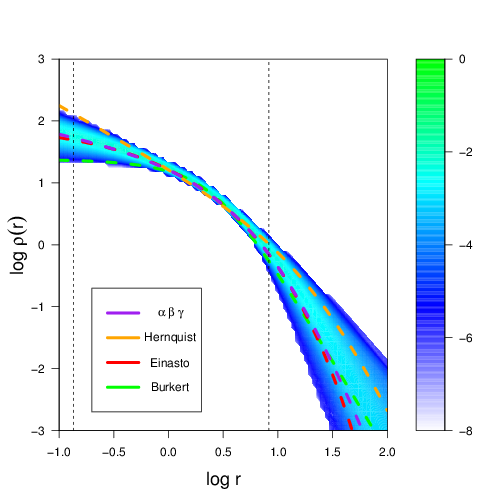}
  \includegraphics[width=\linewidth]{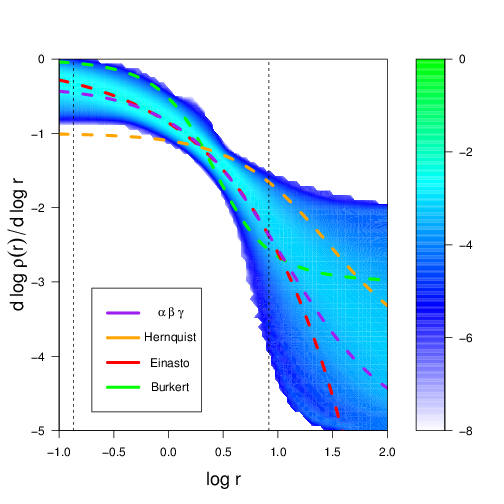}
    \caption{As in Fig.~\ref{fig:lowerrortest} but for DDO 154 data. Here we use free stellar disk parameters $f_{\Upsilon}$ and $f_{R_{\rm d}}$.}
    \label{DDO154test}
\end{figure}

\begin{table*}
\begin{tabular}{ | c | c | c | c | c | c | c | c | c | c |}
  Fitted Profile & $\chi_{\rm red}^2$ & $r_s$ (kpc) & $\rho_s$ ($10^{-3}$M$_\odot$ pc$^{-3}$) & $v_{\rm max}$ (km/s) & n & $\alpha$ & $\beta$ & $\gamma$ & $\gamma_{\rm in}$ \\ \hline
  ($\alpha,\beta,\gamma$) (via MCMC) & $-$ & 10.422 & 37.665 & 48.399 & $-$ & 1.443 & 4.558 & 0.129 & 0.337 \\
  ($\alpha,\beta,\gamma$) (via fitting) & 0.318 & 11.227 & 11.599 & 47.112 & $-$ & 1.115 & 5 & 0.367 & 0.454 \\
 Burkert & 0.55 & 2.648 & 24.171 & 44.395 & $-$ & $-$ & $-$ & $-$ & 0.054 \\
 Einasto & 0.299 & 5.779 & 1.535 & 47.104 & 2.072 & $-$ & $-$ & $-$ & 0.327 \\
 Hernquist & 0.945 & 29.433 & 0.608 & 59.661 & $-$ & 1 & 4 & 1 & 1.014 \\
  \hline
\end{tabular}
\caption{Comparison of the most favoured MCMC bin of DDO 154 with the results of fits of individual halo profiles.}
\label{tblfitsDDO154}
\end{table*}

Fig.~\ref{paramsddo154free} presents the results of our MCMC analysis of DDO154 in the form of histograms of all the model parameters, as well as the physical parameter $\gamma_{\rm in}$. From the latter histogram, we find that the inner log slope of the DDO154 dark matter halo is $\gamma_{\rm in}=0.39\pm0.11$. This result is consistent with the earlier result of~\cite{oh2011} ($\alpha=-0.29\pm0.15$, where $\alpha$ is an inner density slope derived from several data points). As the histogram shows, the data strongly exclude both $\gamma_{\rm in}=1$ and $\gamma_{\rm in}=0$. The $\gamma_{\rm in}$ histogram in Fig.~\ref{paramsddo154free} is similar to those shown in Fig.~\ref{abgtest}, which lends credibility to our MCMC constraints. We note that while the results shown in the Figure are based on runs in which the stellar parameters $f_{\Upsilon}$ and $f_{R_{\rm d}}$ are allowed to vary along the chains, the results obtained when we hold these parameters at their estimated values are very similar. This is consistent with our earlier findings that our analysis is able to return reliable constraints on the halo profiles of LSB galaxies independently of the quality of the constraints on the stellar parameters.  

In order to compare our results for DDO 154 with those of previous studies of this galaxy in a quantitative manner, we have performed fits of individual halo profiles to the observed data in the same way as described in Section~\ref{proffit}. A complete description of all the fits can be found in Table~\ref{tblfitsDDO154}. These fits are also over-plotted on the MCMC data in Fig. \ref{DDO154test}, and their estimates of $\gamma_{\rm in}$ are indicated by the arrows in panel (d) of Fig.~\ref{paramsddo154free}. The $\gamma_{\rm in}$ histogram shows that the Einasto and  ($\alpha,\beta,\gamma$) profile fits yield values of $\gamma_{\rm in}$ which are fully consistent with the distribution of values returned by the MCMC chains. 

However, Table~\ref{tblfitsDDO154} shows that the $\chi^2_{\rm red}$ values for all the individually fitted profiles (including the ($\alpha,\beta,\gamma$) and Einasto profiles) are less than unity. Comparisons between $\chi^2_{\rm red }$ values below unity are meaningless, as this represents consistently approaching data points within the $1\sigma$ error bars. The differences between these low values are essentially a measure of how effectively they model noise. Further, comparison of the $\chi^2_{\rm red }$ obtained from models with different forms and different numbers of fitting parameters is merely indicative of which models are preferred by the data, rather than being formally statistically robust. Our MCMC approach, on the other hand, allows us to compare the relative merit of the full space of ($\alpha,\beta,\gamma$) halo profiles and hence to make meaningful statements about the generic properties of models which are consistent with the observed data. In a  subsequent paper, we will exploit this power to obtain general results for the full set of THINGS galaxies (Hague \& Wilkinson, in prep.).

Early work on the density profile of DDO 154 attempted to compare rotation curves for halo profiles whose parameters were determined by cosmological constraints. \cite{moore1994} excluded the Hernquist profile in favour of a cored profile, but only with a ``cosmological simulation prior'' on the Hernquist parameters. The same prior was later used by \cite{burkert1995} to argue for a ``universal'' dark halo profile. However, given the widely acknowledged limitations of cosmological simulations on kpc scales due to our incomplete understanding of the baryon physics on these scales we have opted not to restrict our parameter space by imposing priors in this manner.

In their analysis, \cite{burkert1995} drew on the work of \cite{flores1994} and concluded that the Burkert profile provides a better fit to the rotation curve of DDO 154 than a $\rho \propto r^{-1}$ cusp. The fact that this profile produced a good quality fit appears inconsistent with our result which excludes a purely cored halo at  approximately the $\sim2.5\sigma$ confidence level. However, the analysis in \cite{burkert1995}  used $r_0 = 2.8$kpc for the Burkert halo and the innermost data point available at the time was at $r = 0.6$kpc. The formula for the log slope of a Burkert profile is
\begin{equation}
{{\rm d} \log \rho \over {\rm d}\log r} = -\left[{r \over r+r_0} + {2r^2 \over r^2 + r_0^2}\right]
\end{equation}
so the shallowest slope within the data range then available was $-0.26$. It is also important to note that this data point has a very large error in $v_{\rm c}$. At the same radius, the $(\alpha,\beta,\gamma)$ model at the peak in the $\gamma(r)$ histogram from our MCMC chains shows a log slope of $-0.58$. \cite{burkert1995}  did not supply a fitting statistic nor errors for the Burkert fit parameters, so it is not possible from this information alone to determine if the result is truly incompatible. However, given the quality of the extant data at that time, it is possible that the uncertainties would be large enough to be consistent with our result. It is also worth noting that in addition to poorer quality velocity data, the stellar data that existed was not as detailed as that obtained from Spitzer and used in \cite{deblok2008}. In particular, from the figures provided by \cite{flores1994} it is clear that the dark matter rotation curve that was needed to reconcile their baryonic and total observed rotation curves was substantially and qualitatively different from that required for the current THINGS/Spitzer derived rotation curves. Nevertheless, we note that the shallowest log slopes from the \cite{burkert1995} and \cite{oh2011} analyses are broadly consistent with that found by our study, with the caveat that they are measured at slightly different radii.

The principle conclusion of \cite{burkert1995} with regard to DDO 154 was that steeply cusped haloes were excluded by the data. Notwithstanding the differences in the observed data used in our study, their conclusion is fully consistent with the $\sim4 \sigma$ exclusion of such haloes by our method.

\section{Conclusion}
\label{sec:conc}

We have presented a new approach to the modelling of disk galaxy rotation curves which uses a Markov Chain Monte Carlo algorithm to explore the parameter space of a very general family of dark matter halo profiles. Through extensive testing on artificial data sets, we have demonstrated that our method can both recover dark matter halo profiles reliably and can provide robust constraints on the distribution of acceptable models in parameter space. More critically, when the observed data are unable to constrain particular parameters, or indeed the entire profile, the output of our algorithm shows this explicitly. This is a useful property, as it means that it does not require pre-judgement or pre-processing to determine whether or not a particular galaxy is amenable to this form of analysis; if a galaxy has too high a surface brightness to allow the dark matter halo parameters to be meaningfully constrained, the algorithm will explicitly show the degeneracy between dark matter and stellar contributions.

In particular, we have shown that the logarithmic slope of the dark matter halo profile at the location of the innermost data point in an observed rotation curve can be robustly constrained in the case of an LSB galaxy. This conclusion is insensitive to moderate over-estimation of the observational errors in the data sets. We have further shown that a reduction in the velocity errors of the data would improve constraints more than an increase in spatial resolution.

We have applied the method to the recently obtained THINGS data set for the LSB galaxy DDO 154, obtaining an estimate of $-0.39 \pm 0.11$ for the logarithmic slope of the dark matter halo at a radius of $0.14$kpc, the radius of the innermost data point in the measured rotation curve. We have compared our MCMC results with those obtained by fitting individual dark matter halo profiles to the data. While the numerical value of our result is consistent with the results of previous studies of this galaxy, the marginalised probability distributions for the parameters that can be produced through MCMC give significantly more information than it is possible to obtain through the fitting of small set of halo profiles. Based on our analysis, it is possible to exclude logarithmic slopes of both 0 and 1 with high significance.

Clearly, our method has the potential to yield interesting results when applied to large samples of galaxy rotation curves. In future work, we intend to apply it to the full set of THINGS galaxies in order to obtain a general picture of the range of halo mass profiles which are consistent with current observed data.

\bibliographystyle{mnras}
\bibliography{citations}

\section{Acknowledgements}

This work made use of THINGS, ``The HI Nearby Galaxy Survey''~\citep{THINGS}. This work is based [in part] on observations made with the Spitzer Space Telescope, which is operated by the Jet Propulsion Laboratory, California Institute of Technology under a contract with NASA. We would like to thank Walter Dehnen, Justin Read, Lia Athanassoula, Rodrigo Ibata for valuable discussions. We wish to thank the anonymous referee for their useful comments. The CosmoMC code was written by Anthony Lewis~\citep{lewis2002}. We would also like to thank Erwin de Blok for providing model data~\citep{deblok2008} in electronic format. MIW acknowledges the Royal Society for support through a University Research Fellowship. PRH acknowledges STFC for financial support. This research used the ALICE High Performance Computing Facility at the University of Leicester. Some resources on ALICE form part of the DiRAC Facility jointly funded by STFC and the Large Facilities Capital Fund of BIS. The project also made extensive use of the Complexity HPC cluster at Leicester which is part of the DiRAC2 national facility.

\appendix
\clearpage
\section{Parameter Transform}
\label{paramtrans}
\begin{figure}
  \includegraphics[width=\linewidth]{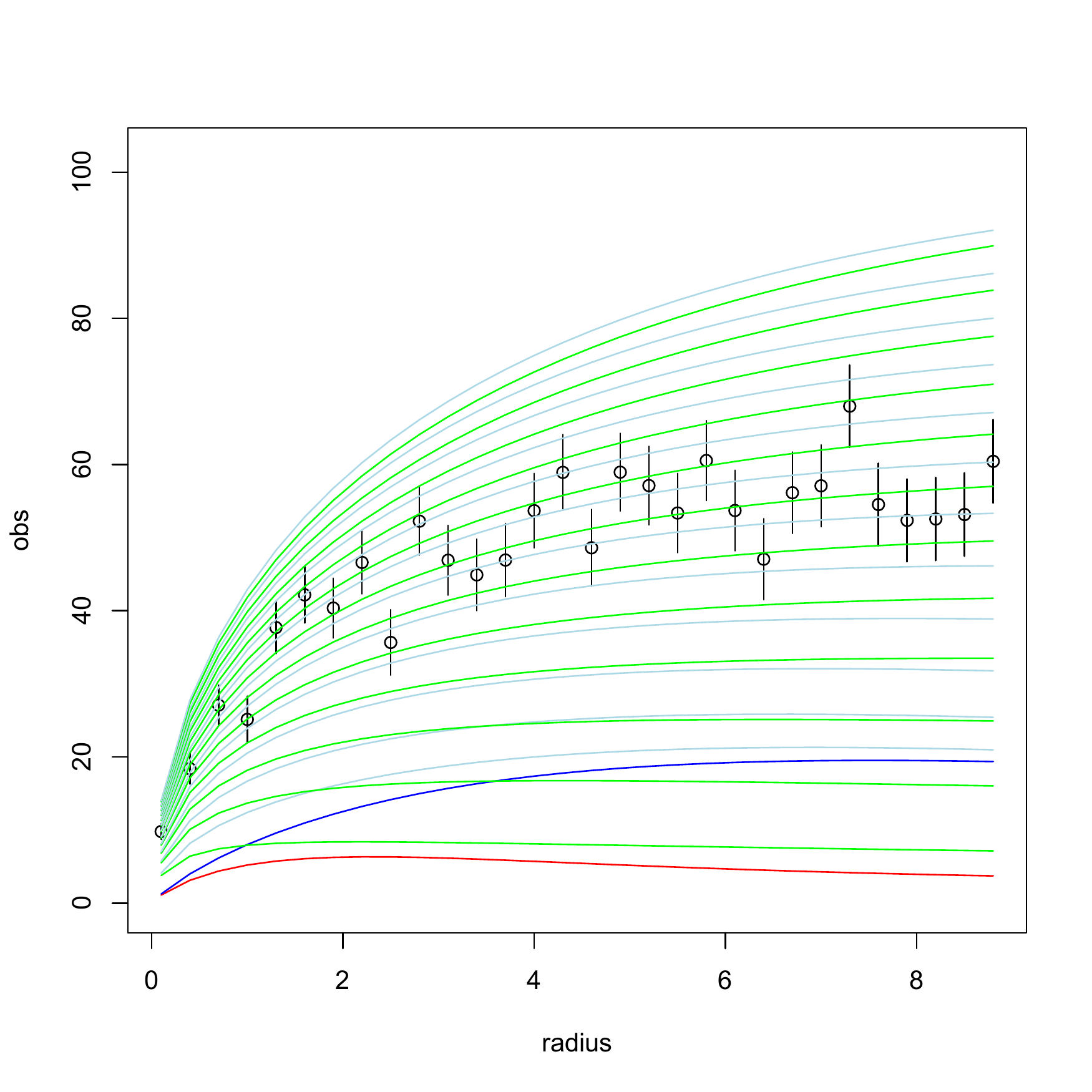}
  \includegraphics[width=\linewidth]{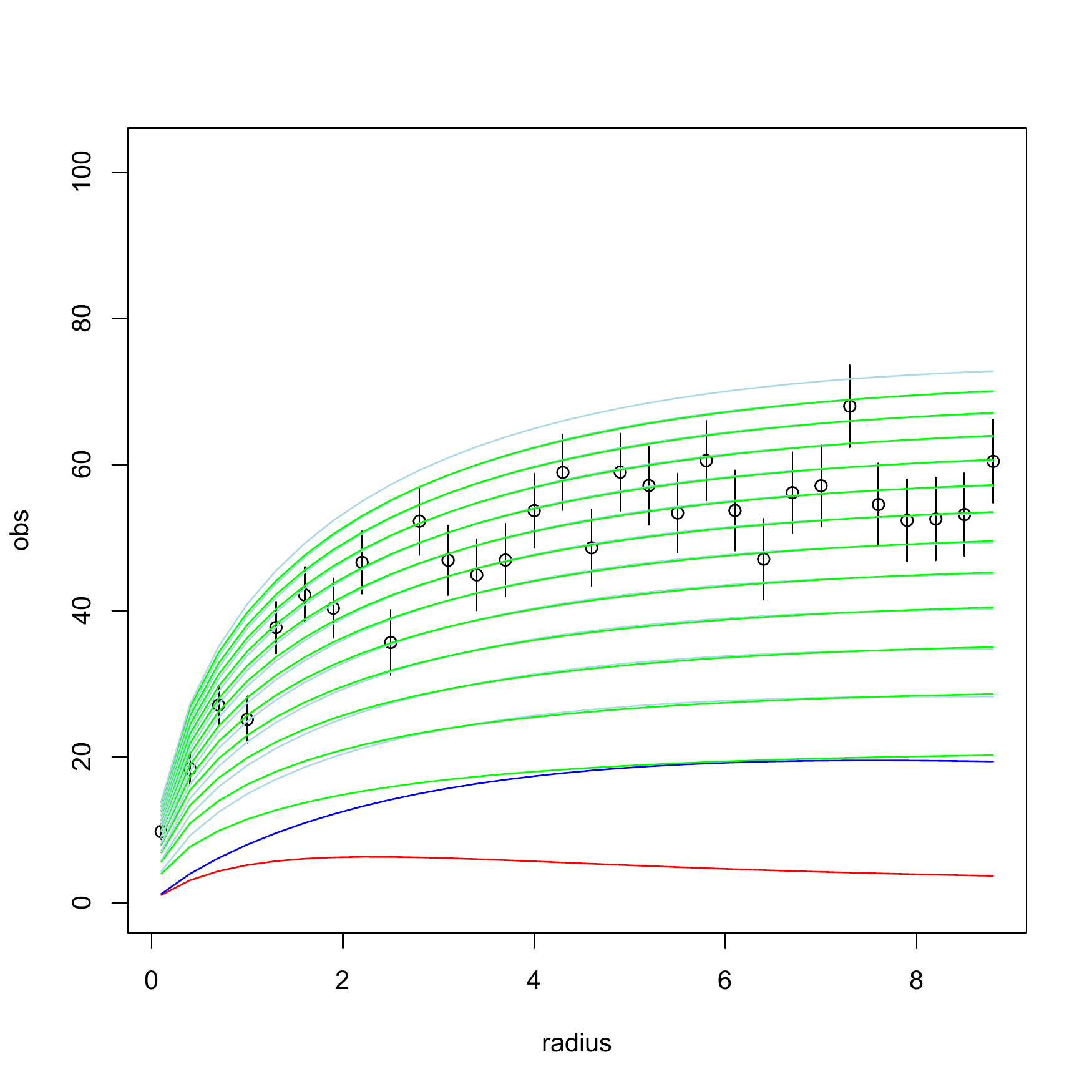}
  \caption{Comparison of the rotation curves obtained from a family of NFW profiles, illustrating the effect of varying $r_{\rm{s}}$ (top panel) and $\rho_{\rm{s}}$ (bottom panel). In each panel, the red line is the stellar contribution to the rotation curve, the blue line is the gas contribution, and the green line is the dark matter halo (with the corresponding total rotation curve shown in light grey). In the upper panel, $r_{\rm{s}}$  varies from 1 kpc (lowest curve) to 12 kpc (highest curve) in steps of 1 kpc while in the lower panel $\rho_{\rm{s}}$ varies from $10^{6}$M$_\odot\,$kpc$^{-3}$ to $1.2\times10^7$M$_\odot\,$kpc$^{-3}$ (highest curve). In the upper panel, $\rho_{\rm{s}} = 6\times10^{6}$M$_\odot\,$kpc$^{-3}$ while in the lower panel $r_{\rm{s}} = 6$kpc.}
    \label{rhordegen}
\end{figure}

Our MCMC chains include the parameter $v_{\rm{max}}$, the peak of the halo rotation curve, from which we must derive the corresponding $\rho_{\rm{s}}$ for the ($\alpha,\beta,\gamma$) profile. Our use of $v_{\rm max}$ in preference to $\rho_{\rm{s}}$ is motivated by the fact that when $\rho_{\rm{s}}$ and $r_{\rm{s}}$ are used as chain parameters, their mutual degeneracy means that both contribute to the amplitude of the halo profile (see Fig.~\ref{rhordegen}). MCMC algorithms are more efficient when the parameters in the chains are chosen to minimise any degeneracies and we found significant improvements in the performance of our algorithm when one parameter scales the halo circular speed curve purely in velocity amplitude, and one scales it purely radially. An additional advantage is that $v_{\rm max}$ is also a physical halo property and hence can be meaningfully compared between galaxies.

For completeness, in this appendix we derive the relation between $v_{\rm{max}}$ and $\rho_{\rm{s}}$ explicitly.

We first define $r_{\rm max}$, the radius at which the circular speed curve of the dark matter halo reaches its maximum value $v_{\rm max}$. Differentiating the expression for the circular speed
\begin{equation}
v_{\rm{circ}}^2(r) = {G M(r) \over r}
\end{equation}
we obtain
\begin{equation}
{d \over dr} v_{\rm circ}^2(r) = 2 v_{\rm circ}(r) G \left(4 \pi r\rho(r) - \displaystyle{M(r) \over r^2}\right)
\end{equation}
whose maximum occurs when
\begin{equation}
\label{vmaxeq}
M(r_{\rm max}) = 4 \pi r_{\rm max}^3\rho(r_{\rm max}) = r_{\rm{max}} {v_{\rm{max}}^2 \over G}
\end{equation}
This is a general result that must be true of any halo profile whose density is only a function of radius. Defining the scaled mass $\tilde{M}(r)$ via
\begin{equation}
M(r) = \rho_{\rm{s}} \tilde{M}(r)
\end{equation}
and rearranging, we obtain the expression to convert between $v_{\rm max}$ and $\rho_{\rm{s}}$
\begin{equation}
\rho_{\rm{s}} = {r_{\rm{max}} v_{\rm max}^2 \over G \tilde{M}(r_{{\rm}max}) }.
\end{equation}
where $r_{\rm{max}}$ is obtained via Eq.~\ref{vmaxeq}, given the values of $v_{\rm max}$ and $r_{\rm s}$.
In the text we combine part of this into the value $\Sigma_{\rm max}$, given by
\begin{equation}
\Sigma_{\rm max} = {r_{\rm{max}} \over \tilde{M}(r_{{\rm}max}) }.
\end{equation}

\end{document}